\documentclass[a4paper,11pt]{article}
\usepackage{jheppub} % for details on the use of the package, please
\usepackage[T1]{fontenc} % if needed
\usepackage{slashed}
\newcommand{\Z}{\mathbb{Z}}

\title{\boldmath Marginal deformations and the Higgs phenomenon
in higher spin AdS$_3$ holography}

\author[a]{Yasuaki Hikida}
\author[b]{Peter B. R\o nne}
\affiliation[a]{Department of Physics, Rikkyo University, \\ 3-34-1 Nishi-Ikebukuro, Toshima, Tokyo 171-8501, Japan}
\affiliation[b]{University of Luxembourg, Mathematics Research Unit, FSTC,
\\ Campus
Kirchberg, 6, rue Coudenhove-Kalergi, L-1359 Luxembourg-Kirchberg, Luxembourg}
\emailAdd{hikida@rikkyo.ac.jp}
\emailAdd{peter.roenne@gmail.com}
\abstract{Recently, a 2d coset model with $\mathcal{N}=3$ superconformal symmetry was proposed to be holographic dual to a higher spin  supergravity on AdS$_3$ and the relation to superstring theory was discussed. However, away from the tensionless limit, there is no higher spin symmetry and the higher spin states are massive.
In this paper, we examine the deformations of the coset model which preserve $\mathcal{N}=3$
superconformal symmetry, but break generic higher spin symmetry.
We focus on double-trace type deformations which are dual to changes of boundary conditions for the bulk matter fields. In the bulk theory, the symmetry breaking will generate mass for the higher spin fields. As a concrete example, we compute the Higgs mass of a spin 2 field both from the bulk and the boundary theory.}

\begin{document}
\maketitle
\flushbottom

\section{Introduction}

Superstring theory contains a plethora of massive excitations and its tensionless limit is believed to be described by higher spin gauge theory.
In fact, it was argued that superstring theory could be described by the broken phase of higher spin gauge theory \cite{Gross:1988ue}.
However, on a flat space-time, the higher spin symmetry is too restrictive, and no-go theorems, e.g., by Weinberg \cite{Weinberg:1964ew},  prohibit a non-trivially interacting theory under some assumptions. Fortunately, these no-go theorems do not apply for a theory with a curved background, and Vasiliev theory constitutes a famous example of a non-trivial higher spin gauge theory defined on AdS space \cite{Vasiliev:2003ev}. Moreover, recent developments of the AdS/CFT correspondence have revealed non-trivial relations between higher spin gauge theory and superstring theory.
The first concrete proposal for this relation (called ABJ triality) was made in \cite{Chang:2012kt} by extending the holographic duality with higher spin gauge theory in \cite{Sezgin:2002rt,Klebanov:2002ja}.
Therefore, it is natural to expect that superstring theory on AdS space can be realized as a broken phase of Vasiliev theory.

In this paper, we would like to examine this relation by studying details of a concrete example.
For this purpose it would be nice to have a more tractable setup than the original ABJ triality. In this case, we want to make use of lower dimensional models.  Generalizing the higher spin AdS/CFT duality in \cite{Gaberdiel:2010pz}, lower dimensional versions of the ABJ triality were proposed in
\cite{Gaberdiel:2013vva,Gaberdiel:2014cha,Gaberdiel:2015mra} using small or large $\mathcal{N}=4$ superconformal symmetry, and also independently in  \cite{Creutzig:2013tja,Creutzig:2014ula} using $\mathcal{N}=3$ superconformal symmetry.
See also \cite{Candu:2013fta,Gaberdiel:2014yla,Beccaria:2014jra,Candu:2014yva} for related works.
The small or large $\mathcal{N}=4$ superconformal symmetry is quite constraining, and one can determine the properties of their triality to a large extent using only the supersymmetry.
Nevertheless, in this paper we consider the case having $\mathcal{N}=3$ superconformal symmetry, the reason being that the analogy with the ABJ triality is more transparent and we expect that the physical intuition can be applied more easily.

In \cite{Creutzig:2014ula} we proposed a holographic duality between a 3d extended Vasiliev theory and a 2d critical level coset model. Utilizing the duality, we discussed the relation to superstring theory.
The higher spin theory is a $\mathbb{Z}_2$ truncation of the $\mathcal{N}=2$ Prokushkin-Vasiliev theory with U$(2M)$ Chan-Paton (CP) factor. If we set $M=2^{n-1}$, then the theory admits $\mathcal{N}=2n+1$ supersymmetry \cite{Prokushkin:1998bq,Henneaux:2012ny}.
The dual CFT is proposed to be the following coset model
\begin{align}
 \frac{\text{su}(N+M)_{N+M} \oplus \text{so}(2NM)_1}{\text{su}(N)_{N + 2M} \oplus \text{u}(1)_{\kappa}}
\label{coset}
\end{align}
with $\kappa = 2NM(N+M)^2$ and several fermions decoupled.
In order to see the relation to the classical Vasiliev theory, we need to take the large $N$ limit while keeping $M$ finite.
Applying the logic of \cite{Chang:2012kt} in this lower-dimensional case, superstring theory should be related to the higher spin theory with some conditions on the CP factor.
We have chosen the U$(M)$ invariant condition on the U$(2M)$ CP factor,
and the dual coset should now be the Grassmannian Kazama-Suzuki model \cite{Kazama:1988uz,Kazama:1988qp}
\begin{align}
 \frac{\text{su}(N+M)_{N+M} \oplus \text{so}(2NM)_1}{\text{su}(N)_{N + 2M} \oplus
\text{su}(M)_{M+2N} \oplus \text{u}(1)_{\kappa}}
\label{KScoset}
\end{align}
with the central charge $c= 3MN/2$.
The most important property of this model is that it admits ${\cal N}=3$ superconformal symmetry as we found in \cite{Creutzig:2014ula}.

We can now examine the superstring theory dual to the Kazama-Suzuki coset \eqref{KScoset} by utilizing the ${\cal N}=3$ superconformal symmetry.
The target space of dual superstring theory should be of the form AdS$_3 \times $M$_7$ where M$_7$ is some 7-dimensional manifold.
In the case with pure NSNS background, a general argument in \cite{Argurio:2000tg} states that the only possible backgrounds are those with M$_7 = $SU(3)$/$U(1), SO(5)$/$SO(3) along with the case M$_7 = ($S$^3 \times $S$^3 \times $S$^1)/\mathbb{Z}_2$ studied in \cite{Yamaguchi:1999gb}.
The BPS spectrum of the superstring theory with
M$_7 = $SU(3)$/$U(1) and SO(5)$/$SO(3) was examined in \cite{Argurio:2000xm},
and the spectrum is consistent with the chiral primaries in the coset model \eqref{KScoset} as argued in \cite{Creutzig:2014ula}, and this will be elaborated on below.%
\footnote{We can check that the BPS spectrum for the case with M$_7 = ($S$^3 \times $S$^3 \times $S$^1)/\mathbb{Z}_2$
 is different.}
Therefore it is expected that the coset model lives on the same moduli space  as
the superstring theory with
M$_7 = $SU(3)$/$U(1) or SO(5)$/$SO(3), even though we know nothing about this moduli space except for a few points.

In this paper, we study the marginal deformations of the coset model \eqref{KScoset} and interpret them in terms of the dual higher spin theory. The two theories stay at the same moduli point which is supposed to correspond to the tensionless limit of superstring theory. Away from this tensionless limit, there is no higher spin symmetry anymore and the higher spin states should become massive. We would like to break the higher spin symmetry quite weakly as in \cite{Maldacena:2011jn}, since otherwise we loose control of the models that was given by the large symmetry algebra. This leads us to consider deformations of the double-trace type which are known to be dual to changes of the boundary conditions for the bulk fields  \cite{Witten:2001ua}.
We examine the marginal deformations of the coset model \eqref{KScoset} while preserving the $
\mathcal{N}=3$ superconformal symmetry. We show that the deformations break a spin 2 symmetry and this, in turn, implies that they break generic higher spin symmetry as well.
In the dual higher spin theory the changes of boundary conditions break higher spin gauge symmetry and the gauge fields become massive due to the symmetry breaking. As a concrete example, we compute the Higgs mass of a spin 2 field both from the bulk and the boundary theories.

It is known that the graviton on AdS space can be massive via loop effects of bulk fields with non-standard boundary conditions \cite{Porrati:2001db,Duff:2004wh}.
In order for a massless gauge field to become massive, it should be swallowing extra degrees of freedom, and in this case they come from bound states of bulk fields.
Moreover, it was pointed out in \cite{Girardello:2002pp} that higher spin gauge fields of 4d Vasiliev theory can be massive in a similar manner with non-standard boundary conditions.
In this paper, we will show how the marginal deformations of the coset model \eqref{KScoset} are mapped to the changes of boundary conditions for the bulk fields of the 3d Vasiliev theory as in \cite{Witten:2001ua}.
From the coset model, we can compute the anomalous dimension of a spin 2 current which is not conserved due to the effects of the marginal deformation as in \cite{Aharony:2006hz,Maldacena:2011jn}. Using the AdS/CFT dictionary, we can read off the Higgs mass of the dual spin 2 field.
From the bulk theory, we compute the Higgs mass directly
by computing the contributions from scalar and fermion loops.
For the scalar loops the results in \cite{Aharony:2006hz, Kiritsis:2006hy} can be used since the set up is found to be the same.
For the fermion loops we have to extend their analysis.

This paper is organized as follows;
In the next section, we review the higher spin gauge theories which are dual to \eqref{coset}
and \eqref{KScoset}.
Primary states of the Kazama-Suzuki coset \eqref{KScoset} are then studied in section \ref{BPS}.
We find new chiral primaries, not considered in \cite{Creutzig:2014ula}
and interpret them in terms of the bulk theory.
In section \ref{marginal} we study marginal deformations of the coset \eqref{KScoset} which preserve $\mathcal{N}=3$ superconformal symmetry.
We show that a spin 2 current is not conserved due to these marginal deformations in section \ref{CFT}, and further we compute the Higgs mass of the dual spin 2 field by using the AdS/CFT dictionary.
Section \ref{higgs} then demonstrates how to reproduce this Higgs mass from the viewpoint of the dual higher spin theory making use of the results in appendices \ref{dtdhol} and \ref{Bulk}.
In section \ref{conclusion} we conclude this paper.
Appendix \ref{N=3} gives a summary of the operator products for generators of the $\mathcal{N}=3$ superconformal algebra.
In appendix \ref{dtdhol} we review the holographic interpretation of the double-trace deformations of a CFT following mainly \cite{Mueck:2002gm,Allais:2010qq,Chang:2012kt}.
Finally, in appendix \ref{Bulk} we review the results for scalar loops in \cite{Aharony:2006hz, Kiritsis:2006hy} and extend them to include the case with fermion loops.

\section{Higher spin gauge theory}
\label{hsgt}

In \cite{Creutzig:2013tja} a higher spin AdS/CFT duality was proposed involving the $\mathcal{N}=2$ supersymmetric version of Prokushkin-Vasiliev theory with $M ' \times M '$ matrix valued fields introduced in \cite{Prokushkin:1998bq}. The duality can be seen as an extension of the $M'=1$ case in \cite{Creutzig:2011fe}. The higher spin theory includes a parameter $\lambda$, which determines the gauge algebra denoted by shs$_{M'}[\lambda]$ as well as the mass of the matter fields. For $\lambda =1/2$, the theory can be truncated consistently by utilizing a $\mathbb{Z}_2$ symmetry, and the resultant theory has $\mathcal{N}=2n+1$ extended supersymmetry for $M '=2^n$ \cite{Prokushkin:1998bq,Henneaux:2012ny,Creutzig:2014ula}.
The higher spin theory is conjectured in \cite{Creutzig:2014ula} to be dual to the coset model presented in \eqref{coset} for $M' = 2M$. For $M'=1$ the duality reduces to the one in \cite{Beccaria:2013wqa} with $\mathcal{N}=1$ supersymmetry when one uses the coset dualities proposed in \cite{Creutzig:2014ula}.%
\footnote{This $\mathcal{N}=1$ duality is different from the one in \cite{Creutzig:2012ar} which has a different truncation of the $\mathcal{N}=2$ supergravity theory.}
The Kazama-Suzuki coset in \eqref{KScoset} is then dual to the higher spin theory, but with a U$(M)$ invariant condition \cite{Creutzig:2014ula}, see \cite{Candu:2013fta}.
In the rest of this section, we review the higher spin theory dual to \eqref{coset} or \eqref{KScoset}.

The supergravity theory dual to \eqref{coset} includes higher spin gauge fields with spin $s=1,2,\cdots$ coupled to matter fields.
The gauge algebra is obtained by a $\mathbb{Z}_2$ truncation of shs$_{M'}[\lambda]$ with $\lambda=1/2$. Let us introduce generators $y_\alpha $ $(\alpha = 1,2)$ and $\hat k$, which satisfy
\begin{align}
 [ y_\alpha , y_\beta ] = 2 i \epsilon_{\alpha \beta} (1 - (1 - 2 \lambda ) \hat k) \, ,
 \quad \hat k^2 = 1\, , \quad \{ \hat k , y_\alpha \} = 0 \, .
 \label{commutator}
\end{align}
We denote the algebra generated by these elements by $sB[\lambda]$. Then the algebra shs$_{M'}[\lambda]$ is obtained by adding the Chan-Paton factors and removing the central element as
\begin{align}
 sB_{M'}[\lambda]  \equiv sB[\lambda] \otimes \mathcal{M}_{M'} = \mathbb{C} \oplus \text{shs}_{M'} [\lambda] \, ,
\end{align}
where $\mathcal{M}_{M'}$ is the $M' \times M'$ matrix algebra and $\mathbb{C}$ is the central element.

At $\lambda = 1/2$, the commutator among $y_\alpha$ in \eqref{commutator} does not involve $\hat k$ any more.
Thus, the algebra can be truncated by requiring the invariance under the $\Z_2$ transformation $\hat k \to - \hat k$, and
the truncated algebra we denote as shs$_{M'}^T[1/2]$. If we set $M' = 2^n$, then the $M' \times M'$ matrix algebra $\mathcal{M}_{M'}$ can be generated by the Clifford elements $\phi^I$ $(I=1,2,\cdots,2n+1)$ with
\begin{align}
 \{ \phi^I , \phi^J \} = 2 \delta^{IJ} \, .
\end{align}
In that case the higher spin algebra shs$_{M'}^T[1/2]$ includes the superalgebra osp$(2n+1|2)$ generated by
\begin{align}
 T_{\alpha \beta} = \{ y_\alpha , y_\beta \} \, , \quad
 Q_\alpha^I = y_\alpha \otimes \phi^I \, , \quad
 M^{IJ} = [\phi^I , \phi^J] \, .
\end{align}
This implies that the theory has $\mathcal{N}=2n+1$ supersymmetry. For the theory dual to \eqref{coset}, we set $M' =2M$. Furthermore  for the theory dual to the Kazama-Suzuki model \eqref{KScoset} we assign the U$(M)$ invariant condition. Even under this condition, the shs$_{2}^T[1/2]$ subalgebra survives and thus the theory still has  $\mathcal{N}=3$ supersymmetry.

Along with higher spin gauge fields, the theory also includes two complex massless scalars conformally coupled to the graviton and two massless Dirac fermions.
It will be convenient to express the $2M \times 2M$ matrix valued fields by $4 \cdot (M \times M)$ matrix fields as
$[\phi_{A \bar B }]^i_{~j}$, $[\tilde \phi_{A \bar B }]^i_{~j}$, $[\psi_{A \bar B }]^i_{~j}$ and $[\tilde \psi_{A \bar B  }]^i_{~j}$
with $A, \bar B  = 1,2$ and $i,j=1,\ldots , M$.  We may represents them by $[\Xi_{A \bar B}]^i_{~j}$.
We are interested in the four single particle fields that are invariant under the U$(M)$ action and which may be expressed as
\begin{align}
 \Xi_{A \bar B } = \text{tr}_M [ \Xi_{A \bar B} ] = [\Xi_{A \bar B}]^i_{~i} \, .
 \label{tracem}
\end{align}
For multi-particle states, the combinations invariant under the U($M$) action are the trace forms $\text{tr}_M [\Xi_{A_1  \bar B_1}] \cdots [\Xi_{A_l  \bar B_l} ]$.
As discussed in appendix \ref{dtdhol},
there are two choices of boundary conditions for the matter fields and
we assign them such that
the dual conformal weights are given by
\begin{align}
 (h , \bar h) = (1/4,1/4) \, , \,   (3/4 ,3/4)\, , \,  (3/4,1/4) \, , \, (1/4,3/4)
\end{align}
for $[\phi_{A \bar B }]^i_{~j}$, $[\tilde \phi_{A \bar B }]^i_{~j}$, $[\psi_{A \bar B }]^i_{~j}$ and $[\tilde \psi_{A \bar B  }]^i_{~j}$, respectively.

The equations of motion for these fields can be found in \cite{Prokushkin:1998bq}, and at the linearized level around the AdS background they are given by (see also \cite{Ammon:2011ua,Creutzig:2012xb,Moradi:2012xd})
\begin{align} \label{fe}
& d A + A  \wedge A = 0 \, , \quad d \bar A + \bar A \wedge \bar A = 0 \, , \\
& d C + A  C - C  \bar A = 0 \, , \qquad
  d \tilde C + \bar A \tilde C - \tilde C  A = 0 \, . \nonumber
\end{align}
Here the 1-forms $A, \bar A$ correspond to higher spin gauge fields and take values in shs$_{2M}^T[1/2]$. Moreover the 0-forms $C, \tilde C$ take values in B$_{2M}[1/2]$ with the invariance under $\hat k \to - \hat k$, and they contain the matter fields. For the theory dual to \eqref{KScoset} we further need to assign the U$(M)$ invariant condition furthermore.

\section{Dual coset model and chiral primaries}
\label{BPS}

The $\mathcal{N}=3$ higher spin gravity with the U$(M)$ invariant condition on the U$(2M)$ CP factor is proposed in \cite{Creutzig:2014ula} to be dual to the Kazama-Suzuki model at the critical level \eqref{KScoset}. The duality holds once we assume a non-diagonal modular
invariant where the factor su$(N+M)_{N+M}$ is expressed by free fermions in the adjoint representation of su$(N+M)$. Generic Kazama-Suzuki models have $\mathcal{N}=2$ supersymmetry, but it was shown  that the critical level model \eqref{KScoset} has enhanced $\mathcal{N}=3$ supersymmetry.
We will deform the coset model while preserving the $\mathcal{N}=3$ superconformal symmetry, so chiral primary states in the undeformed model can be compared with BPS states in superstring theory.
The comparison of the BPS spectrum has been already done in \cite{Creutzig:2014ula}, but we would like to elaborate on the analysis in this section.

\subsection{Primary states}

The generic states of the coset \eqref{KScoset} are labeled by $(\Lambda_{N+M}, \omega ; \Lambda_N , \Lambda_M , m)$ with $\Lambda_L$ denoting a highest weight of su$(L)$. Further, $\omega$ labels the representation of $\text{so}(2NM)_1$, and we will only consider the NS sector given by the sum of the identity $(\omega=0)$ and the vector $(\omega = 2)$ representations. Finally, we have  $m \in \mathbb{Z}_{\kappa}$ giving the u$(1)$ charge.
The states are then obtained by the decomposition
\begin{align}
 \Lambda_{N+M} \otimes \text{NS} = \bigoplus_{\Lambda_N , \Lambda_M ,m }
  (\Lambda_{N+M} ; \Lambda_N , \Lambda_M , m)
  \otimes \Lambda_N \otimes \Lambda_M \otimes m \, .
  \label{deco}
\end{align}
The conformal weight is given by
\begin{align}
 h = n + h^{N+M, N+M}_{\Lambda_{N+M}} + \frac{\omega}{4}
 - h^{N , N+2M}_{\Lambda_N} - h^{M , 2N+M}_{\Lambda_M}
 - h_m \, ,
\end{align}
where
\begin{align}
 h^{L,K}_{\Lambda_L} = \frac{C^L(\Lambda_L)}{K+L} \, , \quad
 h_m = \frac{m^2}{2 \kappa} \, .
\end{align}
We have here denoted the quadratic Casimir of the representation $\Lambda_L$
by $C^L (\Lambda_L)$.
The integer $n$ can be computed by considering how the denominator is embedded in the numerator.
For large $L$, it is convenient to  express the highest weight
$\Lambda_L$ by a set of two Young tableaux $(\Lambda_L^l , \Lambda_L^r)$, and
for large $N$, $m$ is then fixed as (see \cite{Candu:2012jq,Creutzig:2013tja,Candu:2013fta})
\begin{align}
 m = N |\Lambda_{N+M}|_- - (N+M) |\Lambda_N|_-
\end{align}
with the notation $|\Lambda_L|_- = |\Lambda_L^l| - |\Lambda_L^r| $. Here $|\alpha|$ represents the number of boxes in the Young diagram $\alpha$. Since $m$ is uniquely determined, we will suppress it in the following.
We should also take care of field identification among the states \cite{Gepner:1989jq}, but they are irrelevant for large $N$.

In order to construct the model with extended supersymmetry,
we utilize the fact that the factor $\text{su}(N+M)_{N+M}$ can be described by free fermions $\Psi^A$ in the adjoint representation of $\text{su}(N+M)$. We decompose $\text{su}(N+M)$ as follows
\begin{align}
 \text{su}(N+M) = \text{su} (N) \oplus \text{su} (M) \oplus \text{u}(1)
 \oplus (N , \bar{M} ) \oplus (\bar {N} , M)
\end{align}
and we use the same notation as in \cite{Creutzig:2014ula}.
Namely, $\alpha = 1,2 , \ldots , N^2 -1$ and $\rho = 1,2, \ldots , M^2 -1$
are used for the adjoint representations of su$(N)$ and su$(M)$.
Moreover, $a, (\bar a) = 1 ,2 , \ldots N$ and $i ,(\bar \imath) = 1,2, \ldots M$ are for the (anti-)fundamental
representations of su$(N)$ and su$(M)$, which are denoted as $N,M, (\bar N , \bar M)$.
Thus we have the following free fermions
\begin{align}
 \Psi^\alpha \, , \quad \Psi^\rho \, , \quad \Psi^{\text{u}(1)} \, , \quad
\Psi^{(a \bar \imath)} \, , \quad \Psi^{(\bar a i)}
 \, , \quad
 \psi^{(a \bar \imath)} \, , \quad \psi^{(\bar a i)}
\, . \label{fermions}
\end{align}
The last two fermions come from so$(2NM)_1$ in the numerator of \eqref{KScoset}.
The Hilbert space is then generated by these free fermions divided by the denominator of the coset \eqref{KScoset}.
The spectrum may be expressed as
\begin{align}
 {\cal H} = \bigoplus_{\Lambda_{N}, \Lambda_{M}}
 {\cal H}_{ [ \Lambda_{N} , \Lambda_{M} ]}
  \otimes \bar {\cal H}_{ [ \Lambda_{N}^* , \Lambda_{M}^* ]} \, , \quad
 {\cal H}_{ [ \Lambda_{N} , \Lambda_{M} ]} =
 \bigoplus_{\Lambda_{N+M} \in \Omega} (\Lambda_{N+M}; \Lambda_N , \Lambda_M) \, ,
\label{spec}
\end{align}
where we sum over $\Lambda_{N+M} \in \Omega$ satisfying
$\Lambda_{N+M}^l = (\Lambda_{N+M}^r)^t$.
Here $\alpha^t$ represents the transpose of $\alpha$.
See \cite{Beccaria:2013wqa,Creutzig:2014ula} for more details.

\subsection{Chiral primaries}
\label{CP}

In this subsection, we study the chiral primaries of the $\mathcal{N}=3$ Kazama-Suzuki coset \eqref{KScoset}.
Among the  other generators of the $\mathcal{N}=3$ superconformal algebra, the so(3) spin 1 currents are expressed by \cite{Creutzig:2014ula}
\begin{align}
\label{Rcurrents}
& J^3 = \frac{1}{2}  \delta_{a \bar b} \delta_{\bar \imath j} \left( \Psi^{(a \bar \imath )} \Psi^{(\bar b j)} - \psi^{(a \bar \imath )} \psi^{(\bar b j)} \right) \, , \\
& J^+ = \delta_{a \bar b} \delta_{\bar \imath j} \Psi^{(a \bar \imath )} \psi^{(\bar b j )} \, , \quad
 J^- =\delta_{a \bar b} \delta_{\bar \imath j} \psi^{(a \bar \imath )} \Psi^{(\bar b  j )}  \, . \nonumber
\end{align}
The chiral primaries are given by states with $h = q/2$, where
$q$ is the eigenvalue of the zero mode $J^3_0$.
The other states in the same so(3) multiplet may be obtained by the action of $J^-_0$.
See \cite{Miki:1989ri} for the representation theory of the so(3) superconformal algebra.

From the explicit expression of the so(3) spin 1 current $J^3$ in \eqref{Rcurrents}, we can see that $\Psi^\alpha,\Psi^\rho,\Psi^{\text{u}(1)}$ have the charge $q=0$,
$\Psi^{(a \bar \imath )} , \psi^{(\bar a i)}$ have the charge $q=1/2$ and $\Psi^{(\bar a i)},\psi^{(a \bar \imath )} $ have the charge  $q=-1/2$.
Therefore, it is natural to expect that chiral primary states can be constructed by the action of $\Psi^{(a \bar \imath)}$ and $\psi^{(\bar a i)}$.
The fermions $\Psi^{(a \bar \imath)}$ are transforming in the bifundamental representation for $\text{su}(N) \oplus \text{su}(M)$. Utilizing the decomposition \eqref{deco}
we can construct the state as $(\text{adj}; N , \bar{M})$, where adj represents the adjoint representation.
We can check that the state is a chiral primary with $(h,q)=(1/4,1/2)$. In the left-right Young tableaux notation, this state is denoted $((f,f);(f,0),(0,f))$ where $f$ is the (tableaux for the) fundamental representation and $0$ the trivial representation.
The fusions of the chiral primary then lead to other chiral primaries which can be labeled as
$(\Lambda ; (\Lambda^l, 0 ), ( 0 , \Lambda^r ) )$ with $\Lambda = (\Lambda^l ,\Lambda^r)$ satisfying $\Lambda^l = (\Lambda^r)^t$.
Here we have assumed that $M$ is also relatively large, where the comparison to superstring theory is reliable, see \cite{Chang:2012kt,Creutzig:2014ula}.
Similarly $\psi^{(\bar a i)}$ is transforming in the bifundamental representation of $\text{su}(N) \oplus \text{su}(M)$ and is related to $(0; \bar{N} , M)$ or $((0,0);(0,f),(f,0))$ in the left-right tableaux notation and has $\omega=2$.
The state is also a chiral primary with $(h,q)=(1/4,1/2)$.
Taking fusions of the chiral primary, we get other chiral primaries of the form $(0;(0, \Xi^r ), ( \Xi^l , 0) )$ with $\Xi^r = (\Xi^l)^t$. Considering both types of chiral primaries we can generate chiral primaries which are of the form
 $(\Lambda;(\Lambda^l, \Xi^r ), ( \Xi^l , \Lambda^r) )$.

We can confirm that the states  $(\Lambda;(\Lambda^l, \Xi^r ), ( \Xi^l , \Lambda^r ) )$ are chiral primaries by computing the conformal weight $h$ and the so(3) charge $q$.
{}From the construction we can see that $q = \frac12 (|\Lambda^l| + |\Xi^r| )$.
The conformal weight can be computed as
\begin{align}
  h = \frac{|\Lambda^l| + |\Xi^r|}{2}
   - \frac{C^N ((\Lambda^l, \Xi^r )) + C^M (( \Xi^l , \Lambda^r ) )}{2(N+M)}  - \frac{(N+M)^2(|\Lambda^l| - |\Xi^r|)^2 }{4NM(N+M)^2} \, .
\end{align}
For the further computation, it is convenient use that for $\Pi=(\Pi^l,\Pi^r)$ we have \cite{Gross:1993hu}
\begin{align}
C^L (\Pi) = C^L (\Pi^l) + C^L (\Pi^r)  + \frac{|\Pi^l||\Pi^r|}{L}  \, ,
\end{align}
and (see, e.g., appendix A of \cite{Gaberdiel:2011zw})
\begin{align}
  C^L (\alpha) = \frac12 |\alpha| L + \frac12 \left(\sum_i r_i ^2- \sum_j c_j^2 \right)
  - \frac{|\alpha|^2}{2L}
\end{align}
where $r_i$ and $c_j$ are the number of boxes in the $i$-th row and in the $j$-th column, respectively.
In particular we have
\begin{align}
 C^N (\Lambda^l) + C^M ((\Lambda^l)^t)
  = \frac12 |\Lambda^l| (N+M) - \frac{|\Lambda^l|^2}{2N}
  - \frac{|\Lambda^l|^2}{2M} \, .
\end{align}
We can use the above formulas to show that
\begin{align}
 h = \frac{|\Lambda^l| + |\Xi^r|}{4} = \frac{q}{2} \, .
\end{align}
 Therefore the states with $(\Lambda;(\Lambda^l, \Xi^r ), ( \Xi^l , \Lambda^r ) )$ are indeed chiral primaries.

\subsection{Bulk theory interpretation}
\label{bti}

We would now like to interpret
these chiral primaries in terms of the dual higher spin theory.
Let us denote the simplest chiral primaries as
\begin{align}
| c_1 \rangle = |(\text{adj};N , \bar M) \rangle \, , \quad | c_2 \rangle =  |(0;\bar N , M) \rangle \, .
\label{simplest}
\end{align}
Moreover, the simplest anti-chiral primaries are obtained as
$|a_\eta \rangle = J^-_0 |c_\eta\rangle$ with $\eta =1,2$.
In order to compare them with the bulk fields, we have to combine the anti-holomorphic sector as in  \eqref{spec}.  Defining%
\footnote{We need to define the anti-holomorphic currents in a proper way.}
\begin{align}
| \bar c_1 \rangle = | \overline{(\text{adj};N , \bar M)}  \rangle =
 |(\text{adj} ;\bar N , M)   \rangle \, , \quad | \bar c_2 \rangle =  |\overline{(0;\bar N , M) }\rangle =   |(0; N , \bar M) \rangle
\end{align}
and $| \bar a_\eta \rangle$ as in the holomorphic sector, we have the following eight fundamental states
\begin{align}
 | c_\eta \rangle \otimes |\bar c_\eta \rangle \, , \quad
 | c_\eta \rangle \otimes |\bar a_\eta \rangle \, , \quad
 | a_\eta \rangle \otimes |\bar c_\eta \rangle \, , \quad
 | a_\eta \rangle \otimes |\bar a_\eta \rangle \, .
\end{align}
Notice that the same $\eta$ should be used for the holomorphic and anti-holomorphic sectors as in \eqref{spec}.
We would like to identify them as four complex (or eight real) scalars with conformal weight $(h, \bar h)=(1/4,1/4)$;%
\footnote{The action of $J^3_0$ is dual to the action of $ A^3 = \sigma^3 / 2$ from the left hand side to the matrix $[\phi]_{A \bar B}$, see \eqref{fe}. Here we set $\sigma^3 = \left( \begin{smallmatrix} 1 & 0 \\ 0 & -1 \end{smallmatrix} \right)$. \label{foot}}
\begin{align}
 \phi_{11} \, , \quad \phi_{12} \, , \quad \phi_{21} \, , \quad \phi_{22} \, ,
 \label{dualscalars}
\end{align}
respectively,
where we use the notation in \eqref{tracem}.
Fermionic descendants may be obtained by the action of a supercharge $G^3_{-1/2}$  to the above states, see appendix \ref{N=3} for the generators of the $\mathcal{N}=3$ algebra.
These states should be dual to the spin 1/2 fermions  $\psi_{A \bar B}$   with $(h, \bar h)=(3/4,1/4)$.
Similarly we obtain states with $(h, \bar h)=(1/4,3/4)$, which are dual to $\tilde \psi_{A \bar B}$  by the action of $\bar G^3_{-1/2}$.
The application of  $G^3_{-1/2}$ and $\bar G^3_{-1/2}$  to the holomorphic  and anti-holomorphic sectors simultaneously generates states with $(h, \bar h)=(3/4,3/4)$, which are dual to eight real scalars $\tilde \phi_{A \bar B}$ associated with the opposite boundary condition.

Generic chiral primaries may be generated by the fusions of $| c_\eta \rangle$ and
also $|\bar  c_\eta \rangle$ as mentioned above.
In the dual higher spin theory, they should correspond to the U$(M)$ invariants of the products of $[\phi_{11}]^i_{~j}$.
In the case of the ABJ triality, a higher spin field $\varphi^{i}_{~ j}$ with $M \times M$
matrix elements corresponds to a product of bifundamental fields $A^{i}_{~a} B^a_{~j}$ in the ABJ theory \cite{Chang:2012kt}. Here the sum over the $\text{u}(N)$ index $a$ is taken.
A single-string state is known to be dual to a single-trace operator $\text{tr}\, ABAB \cdots AB$, and this should correspond to the singlet product $\text{tr}\,\varphi  \cdots \varphi$.
The corresponding states may be constructed as
$\beta_n ^\dagger |v \rangle = C_n \text{tr}_M [(AB)^n ] | v \rangle$ with a constant $C_n$,
see for instance \cite{Berenstein:2004kk}. Multi-string states correspond to multi-trace operators, and thus the corresponding states may be
expressed as
\begin{align}
 | n_1 , n_2 , \ldots , n_i \rangle
  = \beta_{n_1} ^\dagger \beta_{n_2} ^\dagger \cdots \beta_{n_i} ^\dagger | v \rangle
\end{align}
with $M \geq n_1 \geq n_2 \geq \cdots \geq n_i$.
We would like to identify $n_j$ as the number of boxes in the $j$-th column of a Young diagram. A single string state corresponds to the case where only $n_1$ is non-zero, and this means that the representation of $\text{su}(M)$ should take the form of $[0^{n_1-1},1, 0 , \ldots , 0]$
for single string states.
From the above arguments, we conjecture that single string states correspond to the states with $(\Xi^l , \Lambda^r)$  which are of the form as
$\Xi^l = [0^{\ell-1},1, 0 , \ldots , 0]$ and $\Lambda^r = [0^{p-1},1, 0 , \ldots , 0]$.
 In other words, the chiral primaries corresponding to single-string states have $h = q/2 = (\ell + p)/4$ with non-negative integers $\ell , p$.
The states with other $(\Xi^l , \Lambda^r)$ should correspond to multi-string states.
However, we admit that the map is actually not so precise since it is known that there should be mixing between single-trace and multi-trace operators. So we may use the map just for the purpose of state counting.

Before ending this section, let us comment on a set of important primary states which are not chiral primaries.
Since the coset model \eqref{KScoset} and the higher spin theory stay at the same moduli point, even non-chiral primaries of the coset model have bulk interpretation in the dual higher spin theory.
We can see from \eqref{Rcurrents} that the first type of fermions in \eqref{fermions} have $q=0$, and the products of these fermions
may yield the states of the form
$(\Lambda; \Lambda, 0 )$ with $\Lambda \in \Omega$.
The conformal weight is
\begin{align}
  h = \frac{1}{2(N+M)} (C^{N+M} (\Lambda) - C^N (\Lambda) )
  \sim \frac{M |\Lambda|}{4(N+M)}
\end{align}
for large $N$. Therefore for finite $M$ the conformal weight vanishes and
the corresponding states are the so-called ``light states'' \cite{Gaberdiel:2010pz,Gaberdiel:2011zw}.
In the 't Hooft limit it is argued that they
decouple from the other states and we can consistently remove them from the
spectrum.
We may regard these light states as
duals of non-perturbative geometry dressed with perturbative matter \cite{Castro:2011iw,Tan:2012xi,Datta:2012km,Perlmutter:2012ds,Hikida:2012eu,Campoleoni:2013lma}.
When we discuss the relation to superstring theory, we take $N,M $ large but keep $M/N$ finite as in \cite{Chang:2012kt}.
Within this region these states are no longer light and it is expected that they are not decoupled from the other states.
Since these light states are not chiral primaries, we cannot say anything about them from the dual string viewpoint.

\section{Marginal deformations}
\label{marginal}

We conjecture that
the introduction of finite string tension corresponds to a deformation of the critical coset model \eqref{KScoset}.
In this section we
find the deformations of the coset model \eqref{KScoset} that preserve $\mathcal{N}=3$ superconformal symmetry.
We introduce the deformations of the double-trace type and interpret them
in terms of the dual higher spin theory.

\subsection{Marginal deformations preserving $\mathcal{N}=3$ algebra}

We will now deform the coset model \eqref{KScoset} by adding the deformation term
\begin{align}
 \Delta S = - f \int d^2 w \mathcal{T} (w,\bar w)
 \label{deltas}
\end{align}
to the action.  Let $\mathcal{A}(z)$ denote a generator of the chiral symmetry.
Then the corresponding symmetry is not broken to first order if the following condition is satisfied (see, e.g., \cite{Fredenhagen:2007rx})
\begin{align}
 \oint dw \, \mathcal{T} (w , \bar w) \mathcal{A} (z) = 0 \, .
 \label{defcond}
\end{align}
Here the integral contour is around $w=z$ and with no other insertions within. This condition is equivalent to that the OPE between $\mathcal{T} (w , \bar w)$ and
$\mathcal{A}(z)$ is given by a total derivative of some operator.
For instance, let $\mathcal{A}(z)$ be the energy momentum tensor $T(z)$ and  $\mathcal{T}(w, \bar w)$ a primary operator of the conformal dimension $\Delta$. Since we have
\begin{align}
T(z) \mathcal{T} (w , \bar w ) \sim \frac{\Delta \mathcal{T} (w , \bar w)}{(z-w)^2} + \frac{\partial_w \mathcal{T} (w, \bar w)}{z-w} = \frac{ (\Delta - 1 ) \mathcal{T} (w , \bar w)}{(z-w)^2} + \partial_w \left( \frac{ \mathcal{T} (w, \bar w)}{z-w}  \right)\, ,
\end{align}
the deformation preserves conformal symmetry (or the deformation is marginal) only if $\Delta =1$.

In subsection \ref{CP}
we found several $(h,q)=(1/2,1)$ chiral primaries with $|\Lambda^l| + |\Xi^r| = 2$. Inside the ${\cal N}=3$ multiplet that contains such a chiral primary there is an operator with $(h,q)=(1,0)$,  and we would like to show that these generate exactly marginal deformations preserving the ${\cal N}=3$ supersymmetry.
Let us denote a chiral primary with $(h,q)=(1/2,1)$ by $\Phi_{(1)}$ and the generators of the ${\cal N}=3$ superconformal algebra by
$\{ L_n , G^a_{n+1/2} , J^a_n , \Psi_{n+1/2} \}$
with $a = 1,2,3$ and $n \in \mathbb{Z}$, see appendix \ref{N=3}.
We may construct an ${\cal N}=3$ multiplet from $\Phi_{(1)}$ by the action of the supersymmetry operators
$\{ G^a_{-1/2} , J^a_0 \}$, whose commutation relations are
\begin{align}
  \label{JGcomm}
& [ J^3_0 , J^\pm_0 ] = \pm J^\pm_0 \, , \quad
 [J^+_0 , J^-_0 ] = 2 J^3_0 \, , \\
 &  [J^\pm _0 , G^3_{-1/2}  ] = \mp G^\pm_{-1/2} \, , \quad
  [J^\pm _0 , G^\mp_{-1/2}  ] = \pm 2  G^3_{-1/2} \, , \nonumber
\end{align}
where we have defined
\begin{align}
 J^\pm_n = J^1_n \pm i J^2_n \, , \quad
 G^\pm_{n+1/2} =   G^1_{n+1/2} \pm i G^2_{n+1/2} \, .
\end{align}
In the ${\cal N}=3$ multiplet we define the operators in the spin 1 representation of so(3) algebra  as
\begin{align}
 \Phi_{(1)} \, , \quad \Phi_{(0)} \equiv \frac{1}{\sqrt{2}} J_0^- \Phi_{(1)} \, , \quad \Phi_{(-1)} \equiv \frac12 (J_0^-)^2 \Phi_{(1)} \, ,
\label{defcp}
\end{align}
which in particular satisfy
\begin{align}
 G^+_{-1/2} \Phi_{(1)} = 0 \, , \quad G^-_{-1/2} \Phi_{(-1)} = 0 \, .
 \label{condchiral}
\end{align}
In other words, $\Phi_{(-1)}$ is an anti-chiral primary.

With this notation, we would like to propose that the deformation by the operator
\begin{align}
 {\cal T} = G^{+}_{-1/2} \Phi_{(-1)} - G^{-}_{-1/2} \Phi_{(1)}
 \label{marginalop}
\end{align}
preserves the ${\cal N}=3$ superconformal symmetry.
Here we suppress the anti-holomorphic structure.
What we will explicitly show in the following is that the condition \eqref{defcond} is satisfied by all the generators of the ${\cal N}=3$ superconformal algebra, ensuring preservation of the algebra at linear level.
First of all,
we can show that this operator is singlet under the so$(3)$ algebra as
\begin{align}
& J^+_0 \mathcal{T} = \sqrt{2} G^+_{-1/2} \Phi_{(0)} - 2 G^3_{-1/2} \Phi_{(1)} = 0 \, , \\
 &J^-_0 \mathcal{T} = - 2 G^3_{-1/2} \Phi_{(-1)} - \sqrt{2} G^-_{-1/2} \Phi_{(0)} = 0 \nonumber
\end{align}
from which also follow that $J_0^3 \mathcal{T} = 0$.
In the above equations, we have used  \eqref{JGcomm}, \eqref{defcp} and \eqref{condchiral}.

The ${\cal N}=3$ superconformal algebra includes the ${\cal N}=2$ superconformal algebra as a subalgebra generated by e.g. $G^\pm$.
The proposed form of deformation is known to preserve ${\cal N}=2$ superconformal symmetry, as we will now show explicitly.
We can compute
\begin{align}
& G^\pm_{-1/2} \mathcal{T} = \mp G^\pm_{-1/2} G^\mp_{-1/2} \Phi_{(\pm 1)}
= \mp \{ G^\pm_{-1/2}  , G^\mp_{-1/2} \} \Phi_{(\pm 1)}
  = \mp  4 L_{-1} \Phi_{(\pm 1)} = \mp  4 \partial \Phi_{(\pm  1)} \, , \nonumber \\
&  G^\pm_{1/2} \mathcal{T} = \mp G^\pm_{1/2} G^\mp_{-1/2} \Phi_{(\pm 1)}
= \mp \{ G^\pm_{1/2} , G^\mp_{-1/2} \} \Phi_{(\pm 1)}
  = \mp  ( 4 L_{0} \pm 2 J_0 ) \Phi_{(\pm 1)} = \mp 4 \Phi_{(\pm 1)}  \, ,  \nonumber
\end{align}
where we have used the commutation relations in appendix \ref{N=3}.
We thus have
\begin{align}
 G^\pm (z) {\cal T} (w) \sim \mp  \frac{4 \Phi_{(\pm 1)} (w)}{(z-w)^2} \mp  \frac{ 4 \partial \Phi_{(\pm 1)} (w)}{z-w} = \mp
 \frac{\partial}{\partial w} \left( \frac{ 4\Phi_{(\pm 1)} (w)}{z-w}\right) \, .
\end{align}
Therefore,
the $\mathcal{N}=2$ subalgebra has been shown to be preserved to the first order.
Since the conformal dimension of the deformation operator $\mathcal{T}$ is one, the deformation is marginal to the first order. Actually it was shown that the deformation preserves conformal symmetry to all order of the perturbation (i.e. the deformation is exactly marginal) \cite{Dixon:1987bg}, see also appendix A of \cite{Gukov:2004ym}.

Finally we check the symmetry generators not included in the  $\mathcal{N}=2$ sub-algebra.
For $ \Psi (w)$ we can see
\begin{align}
 \Psi_{1/2} \mathcal{T} = \{ \Psi_{1/2} , G^+_{-1/2} \} \Phi_{(-1)} -
 \{ \Psi_{1/2} , G^-_{-1/2} \} \Phi_{(1)} = J^+_0\Phi_{(-1)} - J^-_0 \Phi_{(1)} = 0 \, ,
\end{align}
which means $\Psi (w) \mathcal{T} (z) \sim 0$.
For $G^3 (w)$ we notice that
\begin{align}
 G^+_{-1/2} (J_0^-)^2 \Phi_{(1)} &= \left( [G^+_{-1/2} , J_0^-] J_0^-
  + J_0^- G^+_{-1/2} J^-_0 \right) \Phi_{(1)}  \\
   & = \left( 2 G^3_{-1/2} J_0^-  + J_0^- [G_{-1/2}^+ , J^-_{0}] \right) \Phi_{(1)}
   = \left( 4 G^3_{-1/2} J_0^-  + 2 [J_0^- , G_{-1/2}^3] \right) \Phi_{(1)} \nonumber \\
     & = \left( 4 G^3_{-1/2} J_0^-  + 2 G^-_{-1/2} \right) \Phi_{(1)} \, ,\nonumber
\end{align}
which leads to
\begin{align}
 {\cal T} =2 \sqrt{2} G^3_{-1/2} \Phi_{(0)} \, .
\end{align}
Since we have
\begin{align}
 & G^3_{-1/2} G^3_{-1/2} \Phi_{(0)} =  L_{-1} \Phi_{(0)}
 = \partial \Phi_{(0)} \, , \quad
 G^{3}_{1/2} G^3_{-1/2} \Phi_{(0)} = 2 L_0 \Phi_{(0)} =  \Phi_{(0)} \, ,
\end{align}
we can show that
\begin{align}
 G^3 (z) {\cal T} (w) \sim \frac{2 \sqrt{2} \Phi_{(0)} (w)}{(z-w)^2} + \frac{2 \sqrt{2} \partial \Phi_{(0)} (w)}{z-w} =
 \frac{\partial}{\partial w} \left( \frac{2 \sqrt{2} \Phi_{(0)} (w)}{z-w}\right) \, .
\end{align}
In this way, we have shown that the deformation by the operator \eqref{marginalop} preserves the $\mathcal{N}=3$ superconformal symmetry to the first order of the perturbation.

\subsection{Double-trace deformations}

As we saw in last subsection, we can construct operators generating deformations preserving the $\mathcal{N}=3$ superconformal symmetry by using chiral primaries $\Phi_{(1)}$ with $(h,q)=(1/2,1)$ and equation \eqref{marginalop}. There are several choices of $\Phi_{(1)}$, but we will be interested in those given by a product of two operators.
There are two types of the simplest chiral primaries $|c_\eta \rangle$ with $(h,q)=(1/4,1/2)$ given in \eqref{simplest} and we now introduce operators $\xi^\eta_{(1/2)}$ creating these states i.e. $|c_\eta \rangle \equiv \xi^\eta_{(1/2)} | 0 \rangle$.
 Combining with the anti-holomorphic sector, we have two operators dual to two real BPS states with alternative quantization $\phi_{11} = \phi_{11}^1 + i \phi_{11}^2$    in \eqref{dualscalars}.
From the simple product of these operators we can construct the chiral primary $\Phi_{(1)}$ with $(h,q)=(1/2,1)$ since the product of chiral primaries does not have any singular terms, as explained in \cite{Lerche:1989uy}.
In the following we consider the case with
\begin{align}
 \Phi_{(1)} =  \xi_{(1/2)}  \xi_{(1/2)} \, ,
 %\quad \xi_{(1/2)}  \equiv \xi^1_{(1/2)} \, ,
 \label{phi1}
\end{align}
where $\xi_{(1/2)}$ is  $\xi^1_{(1/2)} $ or $\xi^2_{(1/2)} $.
We would like to regard
the deformation operator \eqref{marginalop} constructed using \eqref{phi1} as a double-trace deformation, which has a dual interpretation as a change of boundary conditions for bulk fields.
For this purpose we need  to rewrite the deformation operator \eqref{marginalop} in a suitable way.

The marginal deformation $\mathcal{T}$ in \eqref{marginalop} is given by $\Phi_{(\pm 1)}$ with the action of superconformal generators, and we would like to clarify the role of these generators.
We write the doublet $\xi_{(1/2)} , \xi_{(-1/2)} \equiv J_0^- \xi_{(1/2)}$  in the spin $1/2$  spinor representation.
Via the action of the superconformal generators,
we define the following operators with $(h,q)=(3/4,\pm1/2)$ as
\begin{align}
 \xi '_{(1/2)} \equiv \frac{1}{\sqrt{2}} G^+_{-1/2} \xi_{(-1/2)} \, , \quad \xi '_{(-1/2)} \equiv \frac{1}{\sqrt{2}} G^-_{-1/2} \xi_{(1/2)} \, .
\end{align}
The normalization is chosen such that the norm for $\xi '_{(\pm 1/2)}$ is the same as that for $\xi_{(\pm 1/2)}$.%
\footnote{Let us define $| c' \rangle = \xi ' _{(1/2)} | 0\rangle$. Then we find $\langle c' | c' \rangle
 = \frac12 \langle a | G^-_{1/2} G^+_{-1/2} | a \rangle = \frac12 \langle a | (4 L_0 - 2 J_0^3) | a \rangle = \langle a | a \rangle$. }
Combining the anti-holomorphic sector, we introduce
\begin{align} \label{OF}
& {\cal O}^{A \bar B} =  \xi_{(3/2 - A)} \otimes \bar \xi_{(3/2 - \bar B)}  \, , \quad
  {{\cal O}'}^{A \bar B} =  \xi'_{(3/2 - A)} \otimes \bar \xi '_{(3/2 - \bar B)}   \, , \\
 &  {\cal F}^{A \bar B} =  \xi '_{(3/2 - A)} \otimes \bar \xi_{(3/2 - \bar B)}  \, , \quad
  {{\cal F}'}^{A \bar B} =  \xi_{(3/2 - A)} \otimes \bar \xi ' _{(3/2 - \bar B)}  \, .
  \nonumber
\end{align}
As explained in subsection \ref{bti}, ${\cal O}^{A \bar B} $, ${{\cal O}'}^{A \bar B} $ are  dual to scalar fields with the alternative and the standard quantizations $\phi_{A \bar B}$, $\tilde \phi_{A \bar B}$, respectively. Moreover,  ${{\cal F}}^{A \bar B} , {{\cal F}'}^{A \bar B} $  are dual to spin $1/2$ fermionic fields $\psi_{A \bar B}$, $\tilde \psi_{A \bar B}$. Indeed, since we have
\begin{align}
&  \xi '_{(1/2)}  =  \frac{1}{\sqrt{2}}  [G^+_{-1/2} , J^-_0 ] \xi_{(1/2)}
  =  \sqrt{2} G^3_{-1/2} \xi_{(1/2)}   \, , \\
& \xi '_{(-1/2)} = \frac{1}{\sqrt{2}}   [G^-_{-1/2} , J^+_0 ] \xi_{(-1/2)}
  = - \sqrt{2}  G^3_{-1/2} \xi_{(-1/2)} \, ,
\end{align}
we can show that
\begin{align}
& \xi '_{(1/2)} =   \frac{1}{\sqrt{2}}  [J^+_0 , G^-_{-1/2}  ] \xi_{(1/2)} =  J^+ _0 \xi '_{(-1/2)}
  \, ,   \\
 &\xi '_{(-1/2)}  =  -   \frac{1}{\sqrt{2}} [G^+_{-1/2} , J^-_0 ] \xi_{(-1/2)}
  =   J^-_0 \xi '_{(1/2)} \, , \nonumber
\end{align}
which implies that the doublet $\xi'_{(\pm1/2)} $ is in the spin $1/2$ spinor representation.

With the above preparations, we can now rewrite the marginal operator in \eqref{marginalop} as
\begin{align}
{\cal T} = & \tfrac14 (\xi ' _{(1/2)}  \xi_{(-1/2)}  + \xi_{(-1/2)}  \xi ' _{(1/2)}  - \xi ' _{(-1/2)}  \xi_{(1/2)}  - \xi_{(1/2)}  \xi ' _{(-1/2)} ) \nonumber \\
 &\otimes  (\bar \xi ' _{(1/2)}  \bar \xi_{(-1/2)}  + \bar \xi_{(-1/2)} \bar \xi ' _{(1/2)}  - \bar \xi ' _{(-1/2)}  \bar \xi_{(1/2)}  - \bar \xi_{(1/2)} \bar \xi ' _{(-1/2)}  ) \nonumber \\
 = & \tfrac12 \epsilon_{AC} \epsilon_{BD}   \left[ { {\cal O} ' }^{A B} {\cal O}^{CD}  + { {\cal F} ' } ^ {A B} {\cal F}^{CD} \right]  \label{marginalop2}
\end{align}
by combining the anti-holomorphic sector. In the above expression, we have changed the overall normalization such that deformation from equation \eqref{deltas} takes the form
\begin{align}
\label{deforms3}
 \Delta S &= - \frac{f}{2} \epsilon_{AC} \epsilon_{BD}  \int d^2 w   \left[  { {\cal O} ' }^{A B} {\cal O}^{CD}  +  { {\cal F} ' } ^ {A B} {\cal F} ^ {CD} \right] (w,\bar w) \\
 & = - f  \epsilon_{AC} \epsilon_{BD} \int d^2 x   \left[   {{\cal O} ' }^{A B} {\cal O}^{CD}  +  { {\cal F} ' } ^ {A B} {\cal F} ^ {CD} \right] (x_1 ,x_2) \, . \nonumber
 \end{align}
Here we have changed the worldsheet coordinates as $w = x_1 + i x_2, \bar w = x_1 - i x_2$.
As  in \cite{Witten:2001ua} and appendix \ref{dtdhol}, the deformation has a natural interpretation as the change of boundary condition for the dual bulk fields.
This leads us to think of the deformation as double-trace type even there is no trace in the operators $\mathcal{O},\mathcal{O}',\mathcal{F},\mathcal{F}'$.%
\footnote{
This name can be supported by the property of large $N$ factorization for these operators. This property is assumed here, but it can be shown as in \cite{Chang:2011vka}.
}

\section{Symmetry breaking in the coset model}
\label{CFT}

In the previous section, we obtained the operator \eqref{marginalop} which preserves the $\mathcal{N}=3$ superconformal symmetry of the coset model \eqref{KScoset}. It is natural to expect that the deformation breaks higher spin symmetry generically.
In the large $N$ limit, we show that a certain spin 2 current is not conserved anymore. This implies the breaking of generic higher spin symmetry since operator products with the spin 2 current generate other higher spin currents.%
\footnote{It is important to confirm the breaking of generic higher spin symmetry in a direct way.}
The breaking of higher spin gauge symmetry should thus also occur in the dual bulk theory and this will give rise to the Higgs mass of higher spin fields.
We will now calculate the anomaly of the spin 2 current and use this to compute the mass of the dual spin 2 field.

\subsection{Symmetry breaking}

We start from a generic situation with a spin $s$ current ${\cal A}^{(s)}(z)$. Here $s =2,3,4,\ldots$ for bosonic currents and $s=3/2,5/2,\ldots$ for fermionic currents.  The corrections of the chiral symmetry current to the first order in the perturbation can be computed from
 \begin{align}
  \mathcal{A}^{(s)} (z) f \int d^2 w \mathcal{T}(w,\bar w) \, .
  \label{AT}
 \end{align}
 Let us assume ${\cal A}^{(s)}_{r} \mathcal{T} = 0$ for $r > 0$, then the operator product can be expressed as
 \begin{align}
  \mathcal{A}^{(s)} (z)  \mathcal{T} (w , \bar w)
   = \sum_{l=0}^{[s-1]} \frac{1}{(z - w)^{l+1}} (\mathcal{A}^{(s)} _{-s + l + 1} \mathcal{T}) (w,\bar w) \, .
 \end{align}
 Here $[t]$ is the biggest integer number less than $t$. Acting with the derivative $\partial_{\bar z}$, we have
  \begin{align}
 \partial_{\bar z}  \mathcal{A}^{(s)} (z)  \mathcal{T} (w , \bar w)
   = 2 \pi  \sum_{l=0}^{[s-1]} \frac{(-1)^l}{l!} \partial_z^l \delta^{(2)} (z-w) ( \mathcal{A}^{(s)} _{-s + l+1} \mathcal{T}) (w,\bar w) \, ,
    \end{align}
where we have used $\partial_{\bar z} (z - w) ^{-1} = 2 \pi \delta^{(2)} (z-w)$. From \eqref{AT} we can read off the non-conservation of the chiral symmetry current as (see, e.g., \cite{Gaberdiel:2013jpa})
  \begin{align}
 \partial_{\bar z}  \mathcal{A}^{(s)} (z , \bar z)
   = 2 \pi  f \sum_{l=0}^{[s-1]} \frac{(-1)^l}{l!} \partial_z^l ( \mathcal{A}^{(s)} _{-s + l+1} \mathcal{T}) (z,\bar z) \, .
   \label{currentbreaking}
 \end{align}
 When the right hand side vanishes,  \eqref{defcond}  is satisfied and the current is still holomorphic.

Along with the energy momentum tensor $T$, the Kazama-Suzuki coset \eqref{KScoset} would have spin 2 currents $T^a$ $(a=1,2,3)$ in the adjoint representation of so$(3)$.
We then focus on a specific example with a spin 2 current $T^3$, which may be effectively expressed
as a composite operator $T^3 = 2 T J^3_0$ in the large $N$ limit.%
\footnote{We propose this from the fact that there are four  dual spin 2 fields (or gravitons) $[h_{\mu\nu}]_{A \bar B}$ which are U$(M)$ singlets. The trace element should be dual to $T$, while the element  proportional to $\sigma^3$ should be dual to $T^3$. See also footnote \ref{foot}.}
We will see it convenient to use linear combinations
\begin{align}
 T^{11} = \frac12 (T + T^3) = T P_+ \, , \quad
 T^{22} = \frac12 (T - T^3) = TP_- \, , \quad
 P_\pm = \frac12 (1 \pm 2 J_0^3)
 \label{tii}
\end{align}
instead of $T^3$ itself.
The central charges of $T^{11}$ and $T^{22}$ are
\begin{align}
c_1 = c_2 = \frac{c}{2} \, ,
\label{c1c2}
\end{align}
respectively.
Now the deformation operator in \eqref{marginalop2} is written in terms of operators defined in \eqref{OF}, and the deformation operator consists of the following type of terms as
\begin{align}
 \Delta S = - \frac{f}{2} \int d^2 w {\cal K}^{(1)} {\cal K}^{(2)} (w,\bar w) \, .
 \label{deforms2}
 \end{align}
 The operators are
$
 {\cal K}^{(i)} = {\cal O}^{i \bar B},{{\cal O}'}^{i \bar B},{\cal F}^{i\bar B},{{\cal F}'}^{i \bar B}
$ $(\bar B = 1,2)$ and ${\cal K}^{(i)}$
has non-trivial OPE only with $T^{ii}$.
Denoting the conformal weight of ${\cal K}^{(i)}$ by $(h_i , \bar h_i )$,
we have $h_1,h_2 = 1/4,3/4$ with $h_1 + h_2 = 1$, and similarly for $\bar h_i$.
Notice that there are $ n_0 = 4$ terms with bosonic operators and  $ n_{1/2} = 4$ terms with fermionic operators.
We use $T^{ii}$ instead of $T^3$
since
now the system for the bosonic sector can be identified with the one analyzed in \cite{Kiritsis:2006hy,Aharony:2006hz}, and we can utilize their analysis.

With the above definitions get the following OPEs
\begin{align}
  T^{11} (z) {\cal K}^{(1)} {\cal K}^{(2)} (w, \bar w) \sim
   \frac{h_1 {\cal K}^{(1)} {\cal K}^{(2)} (w , \bar w ) }{(z-w)^2} +  \frac{(\partial {\cal K}^{(1)}) {\cal K}^{(2)} (w , \bar w ) }{z-w} \, , \\
  T^{22} (z) {\cal K}^{(1)} {\cal K}^{(2)} (w, \bar w) \sim
   \frac{h_2 {\cal K}^{(1)} {\cal K}^{(2)} (w , \bar w ) }{(z-w)^2} +  \frac{{\cal K}^{(1)}(\partial {\cal K}^{(2)} ) (w , \bar w ) }{z-w} \, .
\end{align}
Using the generic expression in \eqref{currentbreaking},
we can rewrite them in terms of current non-conservation as
\begin{align}
 \bar \partial T^{11} =   \pi f [   (\partial {\cal K}^{(1)}) {\cal K}^{(2)} - h_1 \partial ({\cal K}^{(1)} {\cal K}^{(2)} )] \, , \\
 \bar \partial T^{22} =  \pi f [ {\cal K}^{(1)}(\partial  {\cal K}^{(2)} ) - h_2 \partial ({\cal K}^{(1)} {\cal K}^{(2)} )  ] \, .
\end{align}
Therefore we have
\begin{align}
 \bar \partial T = 0 \, , \quad \bar \partial T^3 = 2 \pi f [ h_2(\partial {\cal K}^{(1)}) {\cal K}^{(2)} - h_1 {\cal K}^{(1)}(\partial  {\cal K}^{(2)} ) ] \, .
 \label{generict3}
\end{align}
The first equation just means that the conformal symmetry is preserved at the first order perturbation as seen in the previous section.
The second equation indicates that the spin 2 current $T^3$ is broken by the marginal deformation \eqref{deforms2}.

\subsection{Higgs mass from the dual CFT}

In order for a higher spin gauge field to acquire a non-zero mass,  it should swallow the degrees of freedom from the Goldstone modes.
Let us consider a generic $d$ dimensional CFT with higher spin symmetry.
Without deformation, higher spin currents
are conserved as $\partial \cdot  J^{(s)}  = 0$.
After the marginal deformations, the higher spin currents are generically no longer conserved and satisfying
\begin{align}
\partial \cdot J^{(s)} = \alpha \mathcal{O}^{(s-1)} \, .
\label{breaking2}
\end{align}
The divergence of currents are related to another set of operators $\mathcal{O}^{(s-1)}$, and this is dual to the phenomena that the massless gauge fields acquire extra degrees of freedom by the Higgs mechanism. In our example, the operator $\mathcal{O}^{(s-1)}$ is given by a double-trace type as in \eqref{generict3}, and this is related to the fact that the higher spin symmetry is  broken only slightly as in \cite{Maldacena:2012sf}.

The Higgs mass of the higher spin field can be read off from the anomalous dimension of the higher spin current $J^{(s)}$.
We denote by $L_{MN}$ the generators of so$(2,d)$ that is the isometry algebra of AdS$_{d+1}$.
A bulk particle can be classified by the representation of the subalgebra $\text{so}(2) \oplus \text{so}(d)$ as $(E_0, \gamma)$.
The second Casimir is (see, e.g., \cite{Vasiliev:2004cm})
\begin{align}
Q = \frac12 L_{MN}^2 = E_0 (E_0 - d) + 2 C_2^d (\gamma) = M^2_\Delta \, ,
\label{2ndCas}
\end{align}
where $E_0$ corresponds to the conformal dimension of dual operator i.e. $E_0 = \Delta$ and $C_2^d (\gamma)$ is the value of the second Casimir for the representation $\gamma$ of so$(d)$.
For instance, $C_2^d = s (d + s -2 )$ for the $s$-th totally symmetric representation.
 A bulk field satisfying its equation of motion is given by the eigenfunction of the Casimir operator $Q$ with eigenvalue $M^2_\Delta$, see also appendix \ref{propagator}.
 The eigenvalue $M^2_\Delta$ is not the mass square for bulk field and the contribution from AdS curvature should be extracted. From the unitarity bound we know that $\Delta = d+s-2$ for the conserved current with spin $s$, and this leads to $M^2_{d+s-2} = (d+s-2)(2s-2)$ for the dual massless higher spin field. Subtracting this value, we have%
\begin{align}
 M_{(s)} ^2 =\Delta (\Delta - d)  - (d+s-2)(s-2) \, .
 \label{mass2}
\end{align}
This is the formula we will use to compute the Higgs mass.

The anomalous dimension $\Delta$ of the higher spin current $J^{(s)}$ may be computed in the following way.
As in appendix A of \cite{Maldacena:2012sf} (and section 2 of \cite{Aharony:2006hz} for $s=2$) we have
\begin{align}
 | \partial \cdot J^{(s)}| ^2 \propto (\Delta -s -d +2 ) \langle J^{(s)} | J^{(s)} \rangle \, .
\end{align}
Since the RHS of \eqref{breaking2} leads to
\begin{align}
 |\alpha \mathcal{O}^{(s-1)}|^2 = \alpha^2  \langle \mathcal{O}^{(s-1)} | \mathcal{O}^{(s-1)} \rangle \, ,
 \label{RHS}
\end{align}
we can obtain $\Delta$ by equating the above two equations.

We apply the above method for $d=2$ and a spin 2 current $T^3$ with
\begin{align}
 (h , \bar h) = \left( \frac{\Delta+2}{ 2} , \frac{ \Delta - 2 }{ 2 } \right ) \, .
\end{align}
Here $\Delta = 2$ for the conserved energy momentum tensor and we expect $\Delta \neq 2$ after the deformation. First we obtain
\begin{align}
 |\bar \partial T^3|^2 = \langle T^3 | \bar L_{-1} \bar L_1 | T^3 \rangle
 = 2  \langle T^3 | \bar L_{0} | T^3 \rangle = 2 \cdot \frac{ \Delta - 2 }{ 2 } \cdot \frac{c}{2} \, ,
\end{align}
where we have used  $\langle T^3 | T^3 \rangle = c/2$.
Computing the right hand side of \eqref{generict3}, we have
\begin{align}
\frac{(\Delta - 2) c}{2} =  ( 2  \pi f )^2  (2 h_1^2 h_2 + 2 h_2^2 h_1) N_1 N_2
\end{align}
with $\langle \mathcal{K}^{(i)}| \mathcal{K}^{(i)} \rangle = N_i$.
We set  $N_1 = N_2 = 1/(2 \pi)$ from \eqref{ba2} and \eqref{baf2},
which are used for the operators dual to the bulk fields with standard kinetic terms.
At the leading order of $f^2$, the formula \eqref{mass2} becomes
$
 M^2_{(2)} = 2 (\Delta - 2)
$.
Since $\mathcal{K}^{(i)}$ could be a spin $1/2$ operator, we have totally the generated mass as
\begin{align}
 M^2_{(2)} =  f^2 \frac{3 }{2c}  ( n_0 + n_{1/2} )
 \label{cftmass}
\end{align}
with $n_0 = n_{1/2} = 4$.

\section{Higgs phenomenon in higher spin theory}
\label{higgs}

In this section we reproduce the mass obtained in \eqref{cftmass} for spin 2 field dual to $T^3$ from the bulk side.
As explained in section \ref{hsgt} the bulk fields are of a $2 M \times 2M$ matrix form, and we consider singlets under the U$(M)$ subgroup denoted by $\Xi_{A \bar B}$ $(A,\bar B=1,2)$ defined in \eqref{tracem}. Thus, we can think of these as being of $2 \times 2$ matrix form and we have additional multiplications with $2 \times 2$ matrix algebra.
We need to know how  spin 2 fields are coupled with scalar fields or spin $1/2$ fermions. The coupling may be read off from the equations of motion in \eqref{fe}. Here we should notice that the Vasiliev theory is given in the frame-like formulation with fields of the form as $A_{\mu}^{~ a_1 \ldots a_{s-1}}$ with one vector index and $s-1$ Lorentz indices. For our purpose, it is convenient to move to the metric-like formulation with the fields having the form $\varphi_{\mu_1 \ldots \mu_s}$ with $s$ being vector indices. We can use the map
\begin{align}
 \varphi_{\mu_1 \ldots \mu_s} = \frac{1}{s} \bar e_{( \mu_1}^{~~a_1} \cdots \bar e_{\mu_{s-1}}^{~~a_{s-1}} A_{\mu_s ) a_1 \ldots a_{s-1}}
\end{align}
with $\bar e_\mu^{~a}$ as the background vielbein at the linearized level.
Without the CP factor, the spin 2 field is just the graviton field $h_{\mu \nu}$ in the metric-like formulation, and we know that the graviton is coupled with matter fields through the bulk energy momentum tensor $\hat T_{\mu \nu}$ as $\kappa h^{\mu \nu} \hat T_{\mu \nu}$. Here we have used $\kappa^2 = 8 \pi G_N$ with the Newton constant $G_N$.

The effects of CP factor can be read off from the equations of motion in \eqref{fe}.
A spin 2 field $[h_{\mu \nu}]_{A \bar B}$ is multiplied to a matter field from right hand side (or from left hand side).
 In the deformation operator in \eqref{deforms2}, the $(11)$ component operator is always paired with $(22)$, and similarly $(12)$ is always paired with $(21)$.
From the rule of multiplication we can see that  $[h_{\mu \nu}]_{11}$ couples only one of the dual paired fields
 $(\Xi_{11}, \Xi_{22})$ or $(\Xi_{12}, \Xi_{21})$ and  $[h_{\mu \nu}]_{22}$ couples the other one. We denote by $\hat T^{(1)}$ and $\hat T^{(2)}$ the bulk energy momentum tensors that couple with $[h_{\mu \nu}]_{11}$ and $[h_{\mu \nu}]_{22}$, respectively.
We can think that $[h_{\mu \nu}]_{11}$ and $[h_{\mu \nu}]_{22}$ as metric fields for two different AdS spaces, and the paired matter fields live in the different spaces. Therefore, we can again identify our set up as the one in \cite{Kiritsis:2006hy,Aharony:2006hz}.
The computation is summarized for the bosonic case and extended to the fermionic case in appendix \ref{Bulk}.

Using the result in \eqref{bulkmass0}, the mass of the spin 2 field dual to $T^3$ is computed from the bulk theory as
\begin{align}
M_{(2)}^2 =  f^2 \frac{G_N}{2} (n_0 + n_{1/2})
\label{bulkmass}
\end{align}
for our setup with $(d, \text{dim}) = (2,2)$.
It is known that the central charge is related to the Newton constant as (see also \eqref{c1c2})
\begin{align}
 \frac{3}{2 G_N} = c_1 = c_2 = \frac{c}{2}
\end{align}
as explained in \cite{Henneaux:2010xg,Campoleoni:2010zq,Campoleoni:2011hg}.
From the relation, we can conclude that the mass from the bulk theory in \eqref{bulkmass} is exactly the same as the mass from the CFT  in \eqref{cftmass}.

\section{Conclusion}
\label{conclusion}

In this paper, we have studied the deformations of the Kazama-Suzuki coset model with $\mathcal{N}=3$ superconformal symmetry in \eqref{KScoset}.
We found that the deformation preserving the $\mathcal{N}=3$ superconformal symmetry should be of the form as in \eqref{marginalop} with a chiral primary $\Phi_{(1)}$ having $(h , q) = (1/2, 1)$. We set the chiral primary to be the product of two chiral primaries $\xi_{(1/2)}$ with $(h,q)=(1/4,1/2)$ as in \eqref{phi1}. Then the deformation can be regarded as being of the double-trace type as is seen in \eqref{marginalop2} or \eqref{deforms3}.

Further, we have shown that in the large $N$ limit a spin 2 symmetry is broken by the deformation and this implies that the generic higher spin symmetry is also broken.  The coset model in \eqref{KScoset} is proposed to be dual to a higher spin gauge theory in \cite{Prokushkin:1998bq}. The double-trace type of deformation is dual to the change of boundary conditions for the bulk scalar and spin $1/2$ fields. This change is also expected to break higher spin gauge symmetry, and the breaking would generate the mass for the higher spin gauge fields. We have computed the Higgs mass of a spin 2 field in \eqref{cftmass} from the coset model and also in \eqref{bulkmass} from the bulk higher spin theory. We can show that the two expressions match by using the parameter mapping of the AdS/CFT correspondence.

An immediate question would be what happens for higher spin fields with $s > 2$. In this paper we have studied a spin 2 field as a simple example and expected that a similar story holds also for generic higher spin fields.
However, this is something we have to confirm. The bulk computation seems to be too complicated to generalize, but the CFT computation looks to be tractable. In fact, we have already obtained partial results on the Higgs masses for generic spin fields at the leading order of $1/c$ using the bulk/boundary correspondence. We would like to report on these results in near future \cite{Creutzig:2015hta}.

We have investigated the holographic duality proposed in \cite{Creutzig:2014ula} because of the connection to superstring theory. Thus the most important task may be to understand the meaning of the marginal deformation for the coset model \eqref{KScoset} in terms of superstring theory. It should be related to the introduction of non-zero string tension, but the precise interpretation is unclear so far.
In order to do so, we need to investigate the moduli space of the dual superstring theory. For example it should be checked whether the superstrings on AdS$_3 \times$M$_7$ with M$_7 = $SU(3)$/$U(1) or SO(5)$/$SO(3) are really related or not. The meaning of parameters for the moduli space should be understood. It should be helpful if we can find the brane construction yielding the string background in the near horizon limit.

In order to obtain the physical meaning in terms of superstring theory, it might be better to utilize
 the other trialities presented in \cite{Chang:2012kt,Gaberdiel:2013vva,Gaberdiel:2014cha,Gaberdiel:2015mra}. The relation to superstring theory is well understood in the ABJ triality of \cite{Chang:2012kt}. Thus it is worth studying the Higgs phenomenon in that case though the computation should be quite involved. For instance, we should study the loop effects of gauge fields with spin $s \geq 1$. In this sense, it might be easier to study the low dimensional holography with $\mathcal{N}=4$ superconformal symmetry in \cite{Gaberdiel:2013vva,Gaberdiel:2014cha,Gaberdiel:2015mra}. However, it is not a simple task to see how higher spin fields are mapped to strings in their holography. Probably it would be useful to examine similarities and discrepancies among the different types of triality.

\subsection*{Acknowledgements}

We are grateful to T. Creutzig for useful discussions.
The work of YH is supported by JSPS KAKENHI Grant Number 24740170. The work of PBR is funded by AFR grant 3971664 from Fonds National de la Recherche, Luxembourg, and partial support by the Internal Research Project GEOMQ11 (Martin Schlichenmaier), University
of Luxembourg, is also acknowledged.

\appendix

\section{$\mathcal{N}=3$ superconformal algebra}
\label{N=3}

The generators of ${\cal N}=3$ superconformal algebra are the energy momentum tensor $T(z)$, the superconformal currents $G^a(z)$, the so$(3)$ currents $J^a(z)$ and a spin $1/2$ fermion $\Psi(z)$, where $a=1,2,3$. The OPEs are $(c=3k)$
\begin{align}
 & T(z) T(w) \sim \frac{c/2}{(z - w)^4} + \frac{2 T(w)}{(z-w)^2} + \frac{\partial T(w)}{z-w } \, , \nonumber \\
 & G^a (z) G^b (w) \sim  \frac{2c/3 \delta^{ab} }{(z-w)^3} + \frac{2  i\epsilon_{abc} J^c (w)}{(z-w)^2}
+ \frac{2 \delta^{ab} T(w) + i\epsilon_{abc} \partial J^c (w)}{z-w}
\, , \\
&  J^a (z) J^b (w) \sim  \frac{k \delta^{ab}}{(z -w)^2} + \frac{i \epsilon_{abc} J^c (w)}{z-w} \, , \qquad
  J^a (z) G^b (w) \sim  \frac{ \delta^{ab} \Psi(w)}{(z -w)^2} + \frac{i \epsilon_{abc} G^c (w)}{z-w}\, ,  \nonumber \\
 & \Psi (z) G^a (w) \sim \frac{ J^a (w)}{z-w} \, , \qquad
 \Psi(z) \Psi(w) \sim \frac{k}{z-w} \, .\nonumber
\end{align}
 The mode expansions of these generators are expressed by $\{ L_n , G^a_{r}, J^a_{n} , \Psi_{r}\}$ with $n \in \mathbb{Z}$ and $r \in \mathbb{Z} + 1/2$. The commutation relations are
\begin{align}
&[L_m ,L_n] = \frac{c}{12} m (m^2 -1) \delta_{m+n} + (m-n) L_{m+n}   \, , \nonumber \\
&\{ G^a_r , G^b_s \} = \frac{c}{3} \left( r^2 - \frac14 \right) \delta^{ab}\delta_{r+s} + 2 \delta^{ab} L_{r+s} + (r-s) i \epsilon_{abc} J^c_{r+s}
 \, , \\
&[J^a _m , J^b_n] = k m \delta^{ab} \delta_{m+n} + i \epsilon_{abc} J^c_{m+n} \, , \quad
[ J^a_m , G^b_r ] =  m \delta^{ab} \Psi_{m+r} + i \epsilon_{abc} G^c_{m+r} \, , \nonumber \\
&\{ \Psi_r , G_s^a \} = J^a_{r +s} \, , \quad
\{ \Psi_r , \Psi_s \} = k \delta_{r+s} \, . \nonumber
\end{align}

\section{Double-trace deformations and holography}
\label{dtdhol}

In this paper, we encounter deformation of the double-trace type
with operators which have scale dimension $\Delta_i$ with $\Delta_1 + \Delta_2 = d$. The operators are dual to fields with the same mass, but with different boundary conditions. This type of marginal deformation was firstly discussed in \cite{Witten:2001ua} and appears also in the context of ABJ triality \cite{Chang:2012kt}.
We first consider the case with bosonic operators $\mathcal{O}_i (x) $. The deformation is given by
\begin{align}
 S ' = - f \int d^d x \mathcal{O}_1 (x) \mathcal{O}_2 (x) \, .
\label{dtd}
\end{align}
Then we move to the case with fermionic operators $\mathcal{F}_i (x)$.
In that case we consider the following deformation
\begin{align}
 S ' = - f \int d^d x ( \bar {\mathcal{F}}_1 (x) \mathcal{F}_2 (x) +   \mathcal{F}_1 (x) \bar{\mathcal{F}}_2 (x) ) \, .
\label{dtdf}
\end{align}
In this appendix we relate these marginal deformations to the changes of boundary conditions for the dual bulk fields. We mainly follow the arguments in \cite{Chang:2012kt}, see also, e.g., \cite{Mueck:2002gm,Allais:2010qq}.

\subsection{Bosonic case}

We use the Poincare coordinates of Euclidean AdS$_{d+1}$, whose metric is
\begin{align}
 ds^2 = \frac{dz^2 + \sum_{i=1}^d dx_i^2 }{z^2} \, .
\end{align}
Here we set the AdS radius equal to one. In these coordinates, the boundary is at $z=0$.
The action for a real scalar propagating on AdS$_{d+1}$ is given by
\begin{align}
 S = \frac12 \int d^{d+1}  x \sqrt{g} \left ( \partial_\mu \phi \partial^\mu \phi + m^2 \phi^2 \right ) \, .
\end{align}
The mass $m^2$ includes the contribution from the coupling with the background curvature.
We consider the case with $- (d^2/4 -1 ) > m^2 > -d^2/4$. The conformal dimensions of dual operators are given by
\begin{align}
  \Delta_\pm = \frac{d}{2}  \pm \xi \, , \quad \xi = \sqrt{\frac{d^2}{4} + m^2} \, .
\end{align}
Near the boundary $z=0$, a general solution to the equation of motion behaves as
\begin{align}
 \phi = \alpha z^{d/2 - \xi} + \frac{\beta}{2 \xi} z^{d/2 + \xi} \, .
 \label{2xi}
\end{align}
We assume the regularity at $z = \infty$, which relates $\alpha$ and $\beta$ as
\begin{align}
 \beta (x) = \int d^d y \,  G^{\Delta_+}_\phi(x-y) \alpha (y)  \, ,
 \label{ba1}
\end{align}
where
\begin{align}
 G^\Delta_\phi (x-y) =  \frac{N^\Delta_\phi}{|x-y|^{2 \Delta}} \, , \quad
 N^\Delta _\phi = \pi^{-d/2} \frac{(2 \Delta - d)\Gamma (\Delta)}{\Gamma (\Delta - d/2)} \, .
 \label{ba2}
\end{align}

Since the metric diverges at $z=0$, we introduce a cut off at $z=\epsilon$.
Then the on-shell action is evaluated over the boundary as\
\begin{align}
S =- \frac{1}{2} \int d^d x \epsilon^{1-d} \phi \partial_z \phi   \, ,
\end{align}
and it diverges as $\epsilon^{-2 \xi}$ at $\epsilon \to 0$.
In order to remove the divergence we introduce the boundary action
\begin{align}
 \delta S = \frac12 \int d^d x \epsilon^{-d}\left( \frac{d}{2} - \xi \right) \phi^2 \, ,
\end{align}
and then we have the finite action as
\begin{align}
S + \delta S =  - \frac{1}{2} \int d^d x\, \alpha (x) \beta (x) \, .
\end{align}
We obtain the standard normalization in the above expression due to the $2 \xi$ factor
in \eqref{2xi}. As in the appendix C of \cite{Chang:2012kt}, we will use an abbreviated notation as $S = - \frac12 \alpha \beta $, which may be written as
\begin{align}
S = -\frac12 \alpha G^{\Delta^+}_\phi \alpha
\label{baction}
\end{align}
using $\beta = G^{\Delta^+}_\phi \alpha$ in \eqref{ba1}.

We consider $\phi_1$ with the alternative quantization and $\phi_2$ with the standard quantization, and $\mathcal{O}_1$ and $\mathcal{O}_2$ as their dual operators. Near $z=0$, we assume the boundary behaviors as
\begin{align}
 \phi_1 = \frac{\alpha_1}{2 \xi} z^{d/2 - \xi} + \beta_1 z^{d/2 + \xi} \, , \quad
 \phi_2 = \alpha_2 z^{d/2 - \xi} + \frac{\beta_2}{2 \xi} z^{d/2 + \xi} \, .
\end{align}
The expectation value of $\mathcal{O}_1$ corresponds to $\alpha_1$, while the expectation value of $\mathcal{O}_2$ corresponds to $\beta_2$.
We are considering the double trace deformation
in \eqref{dtd}. Since \eqref{baction} is written in terms of $\alpha$, we need to perform a Legendre transform for $\alpha_2$. Introducing the sources $J_i$, the boundary action after the deformation may be written as \cite{Chang:2012kt}
\begin{align}
 S = - \frac12 (2 \xi)^{-2} \alpha_1 G_\phi^{\Delta^+}\alpha_1 - \frac12 \alpha_2 G_\phi^{\Delta^+}\alpha_2 + \alpha_2 \beta_2  '
 - J_1 \alpha_1 - J_2  \beta_2  ' - f \alpha_1 \beta_2 ' \, .
\end{align}
On-shell we have $\beta_2 ' = \beta_2$. The two-point functions without the deformation (i.e., with $f=0$) can be computed from the boundary action as%
\footnote{In order to compute the  expression of $(G_\phi^{\Delta^+})^{-1}$, it is convenient to work with the momentum basis by using the formula $\int d^d x \frac{\exp (ik\cdot x)}{|x|^{2 \Delta}} = 2^{d -2 \Delta} \pi^{d/2} \frac{\Gamma(d/2 - \Delta)}{\Gamma (\Delta)} |k|^{2 \Delta - d}$. }
\begin{align}
\langle \mathcal{O}_1 (x) \mathcal{O}_1 (y) \rangle = - (2 \xi)^2 (G_\phi^{\Delta^+})^{-1}
= G^{\Delta^-}_\phi \, , \quad
\langle \mathcal{O}_2 (x) \mathcal{O}_2 (y) \rangle = G_\phi^{\Delta^+} \, .
\end{align}
Examining the equations of motion, we have
\begin{align}
 J_2 = - f \alpha_1 + \alpha_2 \, , \quad J_1 = - \beta_1 - f \beta_2 \, .
\end{align}
Setting $J_1 = J_2 = 0$, we obtain the deformed boundary conditions for the fields $\phi_1$ and $\phi_2$. Rotating the fields, we define
\begin{align}
 \hat \phi_1 = \frac{1}{\sqrt{1 + \tilde f^2}} ( \phi_1 + \tilde f \phi_2) \, , \quad
 \hat \phi_2 = \frac{1}{\sqrt{1 + \tilde f^2}} ( - \tilde f \phi_1 +  \phi_2)
\end{align}
with $\tilde f = 2 \xi f$.
Then the new fields $\hat \phi_i$ have the same boundary condition as $\phi_i$ before the deformation.  Utilizing the new fields $\hat \phi_i$ the two-point functions among $\phi_i$ can be read off as \cite{Aharony:2005sh}
\begin{align} \label{gij}
 G^{ij}_\phi = \frac{1}{1 + \tilde f^2}
 \begin{pmatrix}
 G_\phi^{\Delta^-}  + \tilde f^2 G^{\Delta^+}_\phi &  \tilde f  G_\phi^{\Delta^-}  - \tilde f G^{\Delta^+}_\phi \\
\tilde f  G_\phi^{\Delta^-}  - \tilde f G^{\Delta^+}_\phi  &  G_\phi^{\Delta^+}  + \tilde f^2 G^{ \Delta^- }_\phi
 \end{pmatrix} \, .
\end{align}
Here $G_\phi^\Delta$ is the propagator of scalar field with dual  dimension $\Delta$ before the deformation,
 and its expression is given in \eqref{ba2}.

\subsection{Fermionic case}

We need expressions similar to \eqref{gij} for the the deformation with fermionic operators in \eqref{dtdf} as well. A similar analysis can be found in \cite{Allais:2010qq}.
Let us denote $\Gamma^a$ for Euclidean so$(d+1)$ Gamma matrices with $\{\Gamma^a , \Gamma^b \} = 2 \delta^{ab}$ and set $\Gamma^{z} = \Gamma^{d+1}$.
The action for a Dirac fermion propagating on AdS$_{d+1}$ is
\begin{align}
 S =  \int d^{d+1}  x \sqrt{g}  \bar \psi \left ( \frac12 ( \overrightarrow{ \slashed{\nabla} } - \overleftarrow{ \slashed{\nabla} }) - m \right ) \psi  \, .
\end{align}
We consider $0  \leq m \leq 1/2$. The conformal dimensions of dual operators are
\begin{align}
 \Delta_\pm = d/2 \pm m \, .
\end{align}
Near the boundary $z=0$, a solution to the equation of motion may behave as
\begin{align}
 \psi = \chi z^{d/2 - m} + \zeta z^{d/2 + m}
\end{align}
with
\begin{align}
 \Gamma^z \chi = - \chi \, , \quad \Gamma^z \zeta =  \zeta \, .
 \label{chirality}
\end{align}
Regularity at $z= \infty$ relates $\chi$ and $\zeta$ as
\begin{align}
  \zeta (x) =  \int d^d y G^{\Delta_+}_\psi (x - y) \chi (y) \, ,
  \label{baf}
\end{align}
where
\begin{align}
G^\Delta_\psi = \frac{N^\Delta_\psi  \Gamma \cdot (x - y)}{|x-y|^{2 \Delta + 1}} \, , \quad
N^\Delta_\psi =
\pi^{-d/2} \frac{\Gamma (\Delta + 1/2)}{\Gamma (\Delta + 1/2 - d/2)} \, .
  \label{baf2}
\end{align}
Notice that these equations are consistent with the assignment in \eqref{chirality}.

The on-shell action at $z = \epsilon $ is%
\footnote{Note that the Gamma matrices on AdS$_{d+1}$ are defined as $ \{\gamma^\mu , \gamma^\nu \} = 2 g^{\mu \nu}$. Related to the so$(d+1)$ Gamma matrices, we have, for instance, $\gamma^z = z \Gamma^z$.}
\begin{align}
 S = - \frac12 \int d^d x \epsilon^{-d } \bar \psi \Gamma^z \psi \, ,
\end{align}
which diverges as $\epsilon^{-2m}$ as $\epsilon \to 0$. We introduce the boundary action
\begin{align}
 \delta S = - \frac12 \int d^d x \epsilon^{-d} \bar \psi \psi
\end{align}
as in \cite{Allais:2010qq}, then the on-shell action becomes
\begin{align}
 S + \delta S = - \int d^d x \bar \chi (x) \zeta (x) \, .
\end{align}
This may be written as $S = - \bar \chi G^{\Delta^+}_\psi \chi$ in the abbreviated form from
$\zeta = G^{\Delta^+}_\psi \chi$ in \eqref{baf}.

We consider $\psi_i$ dual to the operators $\mathcal{F}_i$.
Near $z=0$, we assume
\begin{align}
 \psi_i = \chi_i z^{d/2 - m} + \zeta_i z^{d/2 + m} \, , \quad
  \Gamma^z \chi _i = - \chi _i \, , \quad \Gamma^z \zeta _i =  \zeta _i \, ,
\end{align}
where the expectation values of $\mathcal{F}_1$ and $\mathcal{F}_2$ correspond to
$\chi_1$ and $\zeta_2$, respectively.
The deformation is now as in \eqref{dtdf}.
With the sources $\eta_i , \bar \eta_i$,
the  action becomes
\begin{align}
 S = &-  \bar \chi_1 G_\psi^{\Delta^+} \chi_1 - \bar \chi_2 G_\psi^{\Delta^+}\chi_2 + \bar \chi_2 \zeta_2 ' + \bar{\zeta}_2 ' \chi_2   \\
& - \bar \eta_1 \chi_1 - \bar \chi_1 \eta_1 - \bar \eta_2 \zeta_2 ' - \bar \zeta_2 ' \eta_2  - f ( \bar \chi_1 \zeta_2 ' + \bar \zeta_2 ' \chi_1 ) \, , \nonumber
\end{align}
where we have  $\zeta_2 '  = \zeta_2$  and  $\bar \zeta_2 '  = \bar \zeta_2$ on-shell.
The two-point functions before the deformation are
\begin{align}
\langle \mathcal{F}_1 (x) \bar{\mathcal{F}}_1 (y) \rangle = -  (G_\psi^{\Delta^+})^{-1}
= G^{\Delta^-}_\psi \, , \quad
\langle \mathcal{F}_2 (x) \bar{\mathcal{F}}_2 (y) \rangle = G_\psi^{\Delta^+} \, .
\end{align}
Examining the equations of motion, we have
\begin{align}
 \eta_2 = - f \chi_1 + \chi_2 \, , \quad \eta_1 = - \zeta_1 - f \zeta_2
\end{align}
and their barred expressions.
Using the rotated fields
\begin{align}
 \hat \psi_1 = \frac{1}{\sqrt{1 + f^2}} ( \psi_1 + f \psi_2) \, , \quad
 \hat \psi_2 = \frac{1}{\sqrt{1 +  f^2}} ( - f \psi_1 +  \psi_2) \,  ,
\end{align}
 the two-point functions among $\psi_i$ can be read off as
\begin{align}
 G^{ij}_\psi = \frac{1}{1 + f^2}
 \begin{pmatrix}
 G_\psi^{\Delta^-}  +  f^2 G^{\Delta^+}_\psi &   f G_\psi^{\Delta^-}  -  f G^{\Delta^+}_\psi \\
f  G_\psi^{\Delta^-}  -  f G^{\Delta^+}_\psi  &  G_\psi^{\Delta^+}  + f^2 G^{\Delta^-}_\psi
 \end{pmatrix} \, .
\end{align}
Here $G_\psi^\Delta$ is the propagator of the spinor field with dual dimension $\Delta$ before the deformation, and its expression is given in \eqref{baf2}.

\section{Higgs mass from bulk matter loops}
\label{Bulk}

In section \ref{higgs}, we have shown that the computation for the Higgs mass of a spin 2 field can be reduced to the one in \cite{Kiritsis:2006hy,Aharony:2006hz} for the case with scalar loops.
In this appendix we review their analysis and extend it to the case with spin $1/2$ fermion loops.

\subsection{Setup and prescription}

As in  \cite{Kiritsis:2006hy,Aharony:2006hz}
we prepare a product of two $d$ dimensional CFTs with
energy momentum tensors $T^{(1)}$ and $T^{(2)}$.
We only consider the case where their central charges are  equal as $c_1 = c_2$.
The product theory is deformed by the following marginal operator as
\begin{align}
 - f \int d^d x \mathcal{O}^{(1)} \mathcal{O}^{(2)} \, ,
 \label{kadtd}
\end{align}
where the operators $\mathcal{O}^{(1)}$ and $\mathcal{O}^{(2)}$ live in the different CFTs.
We denote the field dual to $\mathcal{O}^{(i)}$ as $\phi^{(i)}$. In this paper, we consider only massless scalars conformally coupled to the graviton, and in our case the dual conformal dimensions are $\Delta^{(1,2)} = \frac{d \pm 1}{2}$ $(\Delta^{(1)} + \Delta^{(2)}= d ) $.
The deformation is dual to the change of boundary conditions for the bulk fields as seen in appendix \ref{dtdhol}, and this would lead to the breaking of higher spin gauge symmetry generically.
We study the mass of bulk spin 2 fields generated due to the symmetry breaking.

We consider spin 2 fields dual to the boundary energy momentum tensors $T^\pm = (T^{(1)} \pm T^{(2)})/\sqrt{2}$.
As discussed in \cite{Porrati:2001db,Duff:2004wh,Aharony:2006hz}, the generated mass can be read off from the two-point function of bulk  energy momentum tensors $\hat T^\pm_{\mu \nu}  = (\hat T^{(1)}_{\mu \nu}  \pm \hat T^{(2)}_{\mu \nu} )/\sqrt{2}$;
\begin{align}
&\langle \hat T^\pm_{\mu \nu} (x)  \hat T^\pm_{\mu ' \nu'} (y) \rangle  \\
& \quad = \frac12 \left(  \langle \hat T^{(1)}_{\mu \nu} (x) \hat T^{(1)} _{\mu ' \nu'} (y) \rangle  \pm  \langle \hat T^{(1)} _{\mu \nu} (x)\hat T^{(2)} _{\mu ' \nu'} (y) \rangle  \pm \langle \hat T^{(2)} _{\mu \nu} (x)\hat T^{(1)} _{\mu ' \nu'} (y) \rangle  +  \langle \hat T^{(2)} _{\mu \nu} (x)\hat T^{(2)} _{\mu ' \nu'} (y) \rangle  \right) \, .
\nonumber
\end{align}
The bulk energy momentum tensor $\hat T^{(i)}_{\mu \nu}$ is written in terms of bilinears of $\phi^{(i)}$ as in \eqref{energy}, and the two-point function $\langle \hat T^{(i)}_{\mu \nu} (x)  \hat T^{(j)} _{\mu ' \nu'} (y) \rangle$ can be computed by using propagators
\begin{align}
G^{ij}_\phi = a^{ij}_{\Delta^-} G^{\Delta^-}_\phi +  a^{ij}_{\Delta^+} G^{\Delta^+}_\phi
\label{adelta}
\end{align}
in \eqref{gij}. Therefore, the two-point function has terms proportional to $(a_{\Delta^+})^2$, $a_{\Delta^+} a_{\Delta^-}$ and $(a_{\Delta^-})^2$. We already know that the mass is not generated without the deformation, and this fact implies that there is no contribution from the terms proportional to $(a_{\Delta^+})^2$ and $(a_{\Delta^-})^2$.
Therefore,  we only need to take care the term proportional to $a_{\Delta^+} a_{\Delta^-}$.  From \eqref{gij} we have
\begin{align}
 a^{11}_{\Delta^+} a^{11}_{\Delta^-} =  a^{22}_{\Delta^+} a^{22}_{\Delta^-} =
  - a^{12}_{\Delta^+} a^{12}_{ \Delta^-} =  - a^{21}_{\Delta^+} a^{21}_{\Delta^-} =  \tilde f^2
\end{align}
up to the order $\tilde f^2$. Therefore, we can conclude that there is no mass generated for the spin 2 field dual to $T^+ = ( T^{(1)} + T^{(2)})/\sqrt{2}$. For the spin 2 field dual to $T^- = ( T^{(1)} - T^{(2)})/\sqrt{2}$, we just need to compute one of the four terms, say  $\langle \hat T^{(1)} _{\mu \nu} (x)  \hat T^{(2)} _{\mu ' \nu'} (y)\rangle$,  and then multiply factor $ - 1/2 \cdot 4 = -2$. This prescription can be found to be the same as that in \cite{Aharony:2006hz}.

We also consider the following deformation
\begin{align}
- f \int d^d x ( \bar{\mathcal{F}}^{(1)} \mathcal{F}^{(2)}  + \mathcal{F}^{(1)} \bar{\mathcal{F}}^{(2)} ) \, ,
\end{align}
where the operators $\mathcal{F}^{(i)}$ are spin $1/2$ spinors.
We consider only massless fermions, and the dual conformal dimensions are  $\Delta^{(1,2)} =d/2$. The arguments in the bosonic case hold also for the fermionic case.
The bulk energy momentum tensor is given in \eqref{energyf} below in this case.
No mass is generated for the spin 2 field dual to $T^+ = (T^{(1)} + T^{(2)})/\sqrt2$ and the mass for the spin 2 field dual to  $T^- = (T^{(1)} - T^{(2)})/\sqrt2$ can be computed from
$\langle \hat T^{(1)} _{\mu \nu} (x)  \hat T^{(2)} _{\mu ' \nu'} (y)\rangle$  with the multiplication of a factor $-2$.

\subsection{Coordinate system and bi-tensors}

We would like to compute the corrections of the mass of a spin 2 field induced by the one-loop effects of  matter fields. As mentioned above, we need to compute  the two-point function $ \langle \hat T_{\mu \nu} (x) \hat T_{\mu ' \nu '} (y) \rangle $ of the bulk energy momentum tensors. We use $\mu , \nu$ and $\mu ' , \nu '$  for tensor indices at $x$ and $y$, respectively. In a maximally symmetric space-time, the two-point function may be decomposed by the following bi-tensors \cite{Allen:1986tt,D'Hoker:1999jc}
\begin{align}
& {\cal I}_1 = g_{\mu \nu} g_{\mu ' \nu '} \, , \quad
 {\cal I}_2 = \hat n_\mu \hat n_\nu \hat n_{\mu '} \hat n_{\nu'} \, , \nonumber \\
& {\cal I}_3 = g_{\mu \mu '} g_{\nu \nu '} + g_{\mu \nu '} g_{\nu \mu '} \, , \quad
{\cal I}_4 = g_{\mu \nu} \hat n_{\mu '} \hat n_{\nu '} + g_{\mu ' \nu '} \hat n_{\mu} \hat n_{\nu} \, , \label{bitensors} \\
&{\cal I}_5 = g_{\mu \mu '} \hat n_{\nu} \hat n_{\nu '} + g_{\mu \nu '} \hat n_{\nu} \hat n_{\mu '} + g_{\nu \nu '} \hat n_{\mu} \hat n_{\mu '} + g_{\nu \mu '} \hat n_{\mu} \hat n_{\nu '} \, .  \nonumber
\end{align}
We use $\hat n_a \equiv \nabla_a \tilde \mu$ as unit vectors tangent to the geodesic from $x$ to $y$, where $\tilde \mu$ is the geodesic distance. Moreover, $g_{\mu \mu '}$ is the parallel propagator defined in \cite{Allen:1985wd}. We can compute quantities involving these objects by making use of the rule in table 1 of \cite{Allen:1985wd} or table 1 of \cite{D'Hoker:1999jc}.

With the above bi-tensors, we
 define the following three traceless bases as
\begin{align} \label{Tbitensor}
& T_1 = \frac{1}{d (d z^2 + 1)} ( {\cal I}_1 + (d+1)^2 {\cal I}_2 - (d+1) {\cal I}_4 )\, , \nonumber \\
& T_2 = - \frac{1}{d} {\cal I}_1 + \frac{d-1}{d} {\cal I}_2 + \frac12 {\cal I}_3 + \frac1d {\cal I}_4 + \frac12 {\cal I}_5 \, , \\
& T_3 = \frac{1}{2z} (4 {\cal I}_2 + {\cal I}_5) \, .   \nonumber
\end{align}
Here $z = - \cosh \tilde \mu$. We choose the above three bases such that the expressions for $d=3$ reduce to those in (22) of \cite{Duff:2004wh}.
Then a transverse and traceless basis may be written in the form of
\begin{align}
 {\cal T} = a_1 (z) (d z^2 + 1) T_1 + a_2 (z) T_2 + a_3 (z) T_3 \, .
\end{align}
Divergence of this basis is computed as
\begin{align}
 \nabla^\mu {\cal T}_{\mu \nu \mu ' \nu '}
 = &- \sqrt{z^2 - 1} \left( (d z^2 + 1) a_1 ' + 2 d z a_1) \hat n \cdot T_1 + a'_2 \hat n \cdot T_2 + a_3 ' \hat n \cdot T_3 \right)  \\
 &+ a_1 (d z^2 + 1) \nabla \cdot T_1 + a_2 \nabla \cdot T_2 + a_3 \nabla \cdot T_3 \, ,\nonumber
 \end{align}
where
\begin{align}
 \hat n \cdot T_1 = \frac{1}{d z^2 + 1} A \, , \quad
 \hat n \cdot T_2 = 0 \, , \quad
 \hat n \cdot T_2 = \frac{1}{2z} B
\end{align}
with
\begin{align}
 A_{\nu \mu ' \nu '} = ( (d+1) \hat n_\nu \hat n _{\mu '} n_{\nu '} - \hat n_{\nu} g_{\mu ' \nu '}) \, , \quad
 B_{\nu \mu ' \nu '} = ( 2 \hat n_\nu \hat n _{\mu '} n_{\nu '}  + g_{\nu \mu '} \hat n_{\nu '} + g_{\nu \nu '} \hat n_{\mu '} ) \, .
\end{align}
We also have
\begin{align}
& \sqrt{z^2 -1} \nabla \cdot T_1 = - \frac{z (1 + 3d + (d^2 - d) z^2)}{(dz^2 + 1)^2} A + \frac{d+1}{d(dz^2 +1)} B \, , \\
 &\sqrt{z^2 - 1} \nabla \cdot T_2 =  \frac{d^2 + d -2}{2d} B \, , \quad
 \sqrt{z^2 - 1} \nabla \cdot T_3 = \frac{1}{z} A - \frac{1 + dz^2}{2z^2} B \, . \nonumber
\end{align}
Assigning $\nabla \cdot {\cal T} = 0$ we have two equations
\begin{align}
 & z(z^2 -1) a_1' = - (d+1) z^2 a_1 + a_3 \, , \\
 & (z^2 -1) z a_3 ' = 2 z^2 \left( \frac{d+1}{d}\right) a_1 + \frac{z^2 (d^2 + d - 2)}{d} a_2 - (1 + dz^2) a_3 \, . \nonumber
\end{align}
We also need ${\cal T}_{(n)}$ for $a_1 = 1/z^n$ with $n=d+1,d+2 , \cdots$, and the explicit expressions can be obtained by solving these equations. For $d=3$ we reproduce the results in appendix B of \cite{Duff:2004wh}.

It will be convenient to use homogeneous coordinates instead of intrinsic coordinates,
see, e.g., \cite{Fronsdal:1978vb,Duff:2004wh}.
The AdS$_{d+1}$ space-time can be described by a hypersurface $X^M X_M = - 1$
$(M = 0, 1, 2, \ldots , d+1)$ in a $d+2$ dimensional space-time, whose metric is given by
 $\eta_{MN} = \text{diag} (- , + , + , \cdots , +, + , -)$.
Here the AdS radius is equal to one as before.
We denote the homogeneous coordinates by $X^M$ and $Y^{M'}$. We use
$G^{MN} (X) = \eta^{MN} + X^M X^N$ as the $d+1$ dimensional metric and also the operator projecting the vector quantities onto the hypersurface. Tensor fields
$h_{MNP \cdots}(X)$ on the hypersurface satisfy $X^M h_{MNP \cdots}(X) = 0$.

We compute the two-point function $\Sigma_{MNM'N'}(X,Y) = \langle \hat T_{MN} (X) \hat T_{M' N'} (Y) \rangle $ of the bulk energy momentum tensors. The geodesic distance $\tilde \mu$ is related as $Z \equiv X \cdot Y = - \cosh \tilde \mu (\equiv z)$.
In order to express the quantity, we can use bi-tensors  \eqref{bitensors} but now in terms of
\begin{align}
& \hat G_{MM'} (X,Y) = G_{MN} (X) \eta^{NN'} G_{N'M'} (Y) = \eta_{MM'}
  + X_M X_{M'} + Y_M Y_{M'} + Z X_M Y_{M'} \, ,\nonumber  \\
 & N_M(X) = \frac{Y_M + ZX_M}{\sqrt{Z^2 - 1}} \, , \quad N_{M'}(Y) = \frac{X_{M'} + ZY_{M'}}{\sqrt{Z^2 - 1}} \, ,
\end{align}
where we should replace as $g_{M M'} = \hat G_{M M'} - (Z+1) N_M N_{M'}$ and $\hat n_M = - N_M$. Instead of ${\cal I}_3$ and ${\cal I}_5$, it can be convenient to use
\begin{align}
& \tilde {\cal I}_3 = \hat G_{MM'} \hat G_{NN'} + \hat G_{MN'} \hat G_{NM'} \, , \\
 & \tilde {\cal I}_5 = \hat G_{MM'} N_N N_{N'} +\hat G_{MN'} N_N N_{M'} +\hat G_{NM'} N_M N_{N'} +\hat G_{NN'} N_M N_{M'} \, . \nonumber
\end{align}
The relation between the two bases is%
\begin{align}
 \tilde {\cal I}_3 = {\cal I}_3 + (Z+1) {\cal I}_5 + 2 (Z+1)^2 {\cal I}_2 \, , \quad
 \tilde {\cal I}_5 = {\cal I}_5 + 4(Z+1) {\cal I}_2 \, .
\end{align}
Using the property, $X^M h_{MNP\cdots}(X) = 0$, we can neglect the terms with $X_M$ or
$Y_{M'}$ in the bi-tensor basis as in (19) of \cite{Duff:2004wh}.

\subsection{The Higgs mass of spin 2 gauge field}
\label{propagator}

We start to compute the explicit expressions for the propagators by generalizing the analysis in \cite{Duff:2004wh} for the case with generic $d$. Using the propagators,
we evaluate the two-point function $ \langle \hat T_{\mu \nu} (x) \hat T_{\mu ' \nu '} (y) \rangle $ by utilizing the Wick contraction, and from it we read off  the corrections to the mass of a spin 2 field due to the scalar and fermion loops. There would also be a contribution from the loop effects of spin 1 gauge field for $d > 2$ as computed in \cite{Duff:2004wh} for $d = 3$.
Here we do not consider this type of effects since the spin 1 gauge field in our $d = 2$ setup is not dynamical.

The second Casimir of the AdS isometry so$(2,d)$ generated by $L_{MN}$ is given by \eqref{2ndCas}, and
the equations of motion for the bulk fields may be expressed as the eigenvalue equations of the
second Casimir.
For a scalar field $\phi (X)$ with $s=0$, we have $L_{MN} = i (X_M \partial_N - X_N \partial_M )$ and the Klein-Gordon equation is
\begin{align}
 \left( \hat N (\hat N + d) - X^2 \partial^2 - E_0  (E_0 - d) \right) \phi (X) = 0
\end{align}
with $\hat N = X \cdot \partial$. The Green's function $\Delta_0(X,Y) = \Delta_0 (Z)$ can be obtained from
\begin{align}
 \left( (1-Z^2) \partial^2_Z - (d+1) Z \partial_Z + E_0 (E_0 - d) \right)  \Delta_0 (Z) = 0 \, ,
\end{align}
where we have used $\partial ^2 = - \partial^2_Z$ and $\hat N = X \cdot \partial = Z \partial_Z$. We set $E_0 = (d \pm 1)/2$ for a conformally coupled massless scalar,
and we use the solution
\begin{align}
  \Delta^{(\alpha)}_0 =  \frac{\Gamma((d+1)/2)}{(d-1) ( - 2\pi)^{(d+1)/2}} \left( \frac{\alpha_+}{(Z +1)^{(d-1)/2}} + \frac{\alpha_-}{( Z - 1)^{(d-1)/2}} \right) \, .
\end{align}
We reproduce the propagator of flat space at the $X = Y$ limit
\begin{align}
   \Delta^{(\alpha)}_0 \sim - \frac{\Gamma((d+1)/2)}{2(d-1) \pi^{(d+1)/2}}  \frac{1}{|X-Y|^{d-1}} \, ,
\end{align}
when we set $\alpha_+ = 1$. We choose the notation such that the expression become the same as the one in \cite{Duff:2004wh} for $d=3$. For the scalar with standard quantization, we should set $\alpha_+ = - \alpha_- = 1$, and for the scalar with alternative quantization, we should set  $\alpha_+ =  \alpha_- = 1$. With these values the relation between $\alpha_\pm$ and $a_{\Delta_\pm}$ in \eqref{adelta} can be found as
\begin{align}
 a_{\Delta_+} = \frac12 (\alpha_+ + \alpha_-) \, , \quad
 a_{\Delta_-} = \frac12 (\alpha_+ - \alpha_-)  \, .
\end{align}

We move to the spin $1/2$ propagator.
As in \cite{Duff:2004wh}  we define $K = \Gamma^{MN} X_M \partial_N$ with
$\{ \Gamma^M , \Gamma^N \} = 2 \eta^{MN}$, which satisfy $K  (K-d) = \hat N (\hat N + d) - X^2 \partial ^2$.
For a spin $1/2 $ spinor, $L_{MN} = i (X_M \partial_N - X_N \partial_M) + \frac{i}{2} \Gamma_{MN}$ and the value of Casimir operator for  a spinor representation is $C_2^d (s) = d(d-1)/16$. Since  $Q=\hat N (\hat N + d) - X^2 \partial^2 + (d+1)(d+2)/8 - K = K (K-d-1)  + (d+1)(d+2)/8 $, we have a factorized relation $(K - 1/2) (K - 1/2 -d) = E_0 (E_0 - d)$ when acting on the spin $1/2$ state $\Psi (X)$.
Thus the Dirac equations are
\begin{align}
 [ K - (E_0 + 1/2)] \Psi (X) = 0 \, , \quad
 [K + (E_0 - 1/2 - d )] \Psi (X) = 0 \, .
\end{align}
We set $E_0 = d/2$ for a massless fermion.
Since we have
\begin{align*}
 [K - (E_0 + 1/2)] [K + (E_0 + 1/2 - d)] = \hat N (\hat N + d) - X^2 - (E_0 + 1/2)(E_0 + 1/2 - d) \, , \\
  [K + (E_0 - 1/2 - d)] [K - (E_0 - 1/2 )] = \hat N (\hat N + d) - X^2 - (E_0 - 1/2)(E_0 - 1/2 - d) \, ,
\end{align*}
the solutions to the Dirac equation may be obtained as
\begin{align}
  \Psi(X) = [K + (E_0 + 1/2 - d)] \Psi_0 \phi (X)  \quad \text{or}
  \quad  \Psi(X) = [K - (E_0 - 1/2 )] \Psi_0 \phi (X)  \, ,
\end{align}
where $E_0^{(0)} = E_0 + 1/2$ for the first scalar and $E_0^{(0)} = E_0 - 1/2$ for the second scalar.
Moreover, $\Psi_0$ is a constant spinor. With this expression we can derive the fermion propagator from the scalar one.
For $E_0 = d/2$, the propagator for Dirac fermion can be written as
\begin{align}
  \Delta^{(\alpha)}_{1/2} = \frac{\Gamma((d+1)/2)}{2( - 2 \pi)^{(d+1)/2}}\left( \frac{\alpha_+ \Gamma^M (X_M - Y_M)}{(Z +1)^{(d+1)/2}} + \frac{\alpha_- \Gamma^M (X_M - Y_M)}{(Z-1)^{(d+1)/2}} \right) \, .
\end{align}
As before we reproduce the propagator of flat space in the $X = Y$ limit
\begin{align}
   \Delta^{(\alpha)}_{1/2} \sim  \frac{\Gamma((d+1)/2)}{2\pi^{(d+1)/2}}  \frac{\Gamma_M (X^M - Y^M)}{|X-Y|^{d+1}}  \sim \Gamma^M \partial_M \left ( - \frac{\Gamma((d+1)/2)}{2(d-1)\pi^{(d+1)/2}}  \frac{1}{|X-Y|^{d-1}}  \right)  \, , \nonumber
\end{align}
when we set $\alpha_+ = 1$.

With the help of propagators obtained above, we compute the two-point function of the bulk energy momentum tensors.
The energy momentum tensor for a massless conformally coupled scalar is
\begin{align}
 \hat T_{\mu \nu}
 = \frac{d + 1}{2d} \partial_\mu \phi \partial_\nu \phi - \frac{d-1}{2d}
  \phi \nabla_\mu \partial_\nu \phi - g_{\mu \nu} \left( \frac{1}{2d} ( \partial \phi )^2 + \frac{(d-1)^2}{8d} \phi ^2 \right) \, .
  \label{energy}
\end{align}
Here we have used the equation of motion for $\phi$ since we neglect the contact terms.
The two-point function can be computed by applying Wick contractions as
\begin{align}
\langle
 \hat T_{\mu \nu} (x) \hat T_{\mu ' \nu '} (y) \rangle
 =&  \left( \frac{ \Gamma((d+1)/2) ^2 }{4 d  ( - 2\pi)^{d+1}}\right)
 \left( \frac{\alpha_+^2  \left( (1 + d Z^2) T_1 + (1+d) (T_2 + Z T_3) \right)}{(Z+1)^{d+1}} \right. \\ &+
\left. \frac{\alpha_-^2 \left( (1 + d Z^2) T_1 + (1+d) (T_2 - Z T_3) \right)}{(Z-1)^{d+1}} \right) \, , \nonumber
\end{align}
where the bases $T_i$ $(i=1,2,3)$ are defined in \eqref{Tbitensor}.
In the above expression, we have ignored  the term proportional to $\alpha_+ \alpha_-$,
since they are irrelevant as mentioned above.

The energy momentum tensor for a massless Dirac fermion is
\begin{align}
 \hat T_{\mu \nu}
 = \frac12 \bar{\psi} \gamma_{(\mu} ( \overrightarrow{\nabla}_\nu - \overleftarrow{\nabla}_{\nu )} ) \psi \, ,
 \label{energyf}
\end{align}
where we have used the equation of motion. The two-point function can be computed as
\begin{align}
\langle
 \hat T_{\mu \nu} (x) \hat T_{\mu ' \nu '} (y) \rangle
 =&  \left( \frac{ \Gamma((d+1)/2) ^2  \text{dim}}{8  ( - 2\pi)^{d+1}}\right)
 \left( \frac{\alpha_+^2  \left( (1 + d Z^2) T_1 + (1+d) (T_2 + Z T_3) \right)}{(Z+1)^{d+1}} \right. \nonumber \\ &+
\left. \frac{\alpha_-^2 \left( (1 + d Z^2) T_1 + (1+d) (T_2 - Z T_3) \right)}{(Z-1)^{d+1}} \right) \, .
\end{align}
The dimension of gamma matrices is denoted by dim.
The contribution from a massless Dirac fermion is $(d \,  \text{dim} )/ 2$ times that from a massless scalar.
We are interested in the case with $(d, \text{dim}) = (2,2)$.
Since the contribution from a Majorana fermion is half of that from Dirac fermion,
the corrections from a massless scalar and a massless Majorana spinor are the same as expected from the dual CFT point of view.

As explained in \cite{Porrati:2001db,Duff:2004wh} we can read off the mass of a spin 2 field from the term proportional to the exchange of  a massive spin 1 field
\begin{align}
 \Pi_{\mu \nu \mu' \nu '} = - 2 \nabla_\mu \nabla_{\mu '} D_{\nu \nu ' } \, ,
 \end{align}
 where the symmetrization of the indices $(\mu \nu) $ and $(\mu ' \nu ')$ is implicitly assumed. Here $D_{\nu \nu ' } $ is the massive spin 1 propagator. As argued in section 2.2 of \cite{Aharony:2006hz} the conformal dimension of the dual operator is $E_0 = d+1$. The second Casimir for a vector representation of so$(d)$ is $C_2^d (v) = (d-1)/2$, and thus we have $Q =  2d$ on the massive spin 1 state $A_\mu (X)$. The propagator for a spin 1 field with $M_{(1)}^2 = 2d$ can be found in \cite{Allen:1985wd}. With that expression we find
 \begin{align}
  \Pi_{\mu \nu \mu' \nu '} = &\frac{\Gamma((d+3)/2) Z}{(-1)^{d}d \pi^{(d+1)/2} (Z^2-1)^{(d+3)/2}} \nonumber \\
 & \times
  \left((d+2) T_1 \left(d Z^2+1\right)+2 T_2 -(d+2) T_3 \left(Z^2+1\right)\right) \, .
 \end{align}
For large $-Z$, we may expand the expression in terms of ${\cal T}_{(n)}$ introduced above as
\begin{align}
  \Pi_{\mu \nu \mu' \nu '} = \frac{\Gamma((d+3)/2)  (d+2)}{(-1)^{d}d \pi^{(d+1)/2} }
   {\cal T}_{(d+2)} + \cdots \, .
\end{align}
Denoting the numbers of real scalar and Majorana spin 1/2 spinors by respectively $n_0$ and $n_{1/2}$, we can
expand the  self energy of a spin 2 field for large $-Z$ as
\begin{align}
 \Sigma_{\mu \nu \mu ' \nu '} (x,y)&= 8 \pi G_N \langle \hat T_{\mu \nu} (x) \hat T_{\mu ' \nu '} (y) \rangle  \\
 & =  8 \pi G_N  (\alpha_+^2 - \alpha^2_-)\left(n_0 + \frac{d \, \text{dim}}{4} n_{1/2}\right) \left(  -\frac{ \Gamma((d+1)/2) ^2 (d+1) }{4 d  ( - 2\pi)^{d+1}}\right)   {\cal T}_{(d+2)} + \cdots \, , \nonumber
\end{align}
where we have considered only the term proportional to that for the spin 1 exchange.
As explained above we should set $\alpha_+^2 - \alpha_-^2 = 4 \alpha_{\Delta_+} \alpha_{\Delta_-} = - 4 \tilde f ^2 = - 4 f ^2 $ and multiply $-2$ to obtain the mass of
the spin 2 field. The final result is
\begin{align}
 M_{(2)}^2 = 64 \pi G_N  f^2 \left(n_0 + \frac{d \, \text{dim}}{4} n_{1/2}\right)
 \frac{\Gamma((d+1)/2)}{ 2 (d+2)( 4\pi)^{(d+1)/2}} \, .
 \label{bulkmass0}
\end{align}

%\bibliographystyle{JHEP}
%\bibliography{AdS3}

\providecommand{\href}[2]{#2}\begingroup\raggedright\endgroup

\end{document}